%% file: divsim.tex
\documentclass{article}
\usepackage[final]{colm2026_conference}

\usepackage{microtype}
\usepackage{hyperref}
\usepackage{url}
\usepackage{booktabs}
\usepackage{amsmath}
\usepackage{graphicx}
\usepackage{xcolor}
\usepackage{enumitem}
\usepackage{subcaption}
\usepackage{algorithm}
\usepackage{algorithmic}
\usepackage{latexsym}
\usepackage[T1]{fontenc}
\usepackage[utf8]{inputenc}
\usepackage{lineno}

\definecolor{darkblue}{rgb}{0, 0, 0.5}
\hypersetup{colorlinks=true, citecolor=darkblue, linkcolor=darkblue, urlcolor=darkblue}

\title{Disentangling Models from Personas in Heterogeneous LLM Simulations}

\author{Dani Roytburg \\
  Carnegie Mellon University \\
  \texttt{droytbur@andrew.cmu.edu} \\\And
  Daphne Ippolito \\
  Carnegie Mellon University \\
  \texttt{daphnei@andrew.cmu.edu} \\}

\definecolor{cOSS}{HTML}{d62728}\definecolor{cQWE}{HTML}{1f77b4}\definecolor{cMAG}{HTML}{9467bd}\definecolor{cGLM}{HTML}{2ca02c}
\begin{document}
\maketitle

\ifcolmsubmission
\linenumbers
\fi

\begin{abstract}
Multi-agent simulations with large language models (LLMs) often operate networks of agents with a single base model.
This overlooks the inter-model effects which may dominate engagement dynamics in real-world deployments.
To show this, we simulate a heterogeneous social network powered by several different base models and show that the amount of engagement an agent receives depends more on its base model than on its assigned persona.
The attraction or repulsion effects of a base model strengthen dramatically when more models are added in the mix, suggesting that networks dynamics may converge to base model effects at scale.
To help explain this effect, we conduct a series of content-mediating analyses, showing the predictability of base models across contexts as well as the relationship between a model's lexical patterns and an engagement-maximizing style.
In light of recent developments in mass multi-agent interaction, this work underscores the relevance of \textit{heterogeneous} compositions in driving the outcomes of those networks.
\end{abstract}

%

\section{Introduction}

As large language models begin to power large simulations of multi-agent networks, interest has grown in the properties of the ``machine culture'' \citep{brinkmannMachineCulture2023, zhouSOTOPIAInteractiveEvaluation2023} they foster through collective interaction \citep{ferrarottiGenerativeAICollective2026, hashemiEmpiricalStudyCollective2026}. This motivates the study of emergent properties like cooperation \citep{willisEvaluatingCollectiveBehaviour2026}, bias \citep{mehdizadehHomophilyinducedEmergenceBiased2025, piaoEmergenceHumanlikePolarization2025}, and misalignment \citep{mouIndividualSocietySurvey2024, huang2025dynamicsmultiagentllmcommunities}.
The salience of LLM social networks such as Moltbook \citep{holtzAnatomyMoltbookSocial2026} and Chirper \citep{zhuCharacterizingLLMdrivenSocial2025} makes such questions relevant to the broader public. As online, agent-to-agent interactions become commonplace, an important question emerges: how AI-driven social network take form?

\begin{figure}[!t]
\centering
\begin{minipage}[c]{0.50\textwidth}
  \centering
  \includegraphics[width=\linewidth]{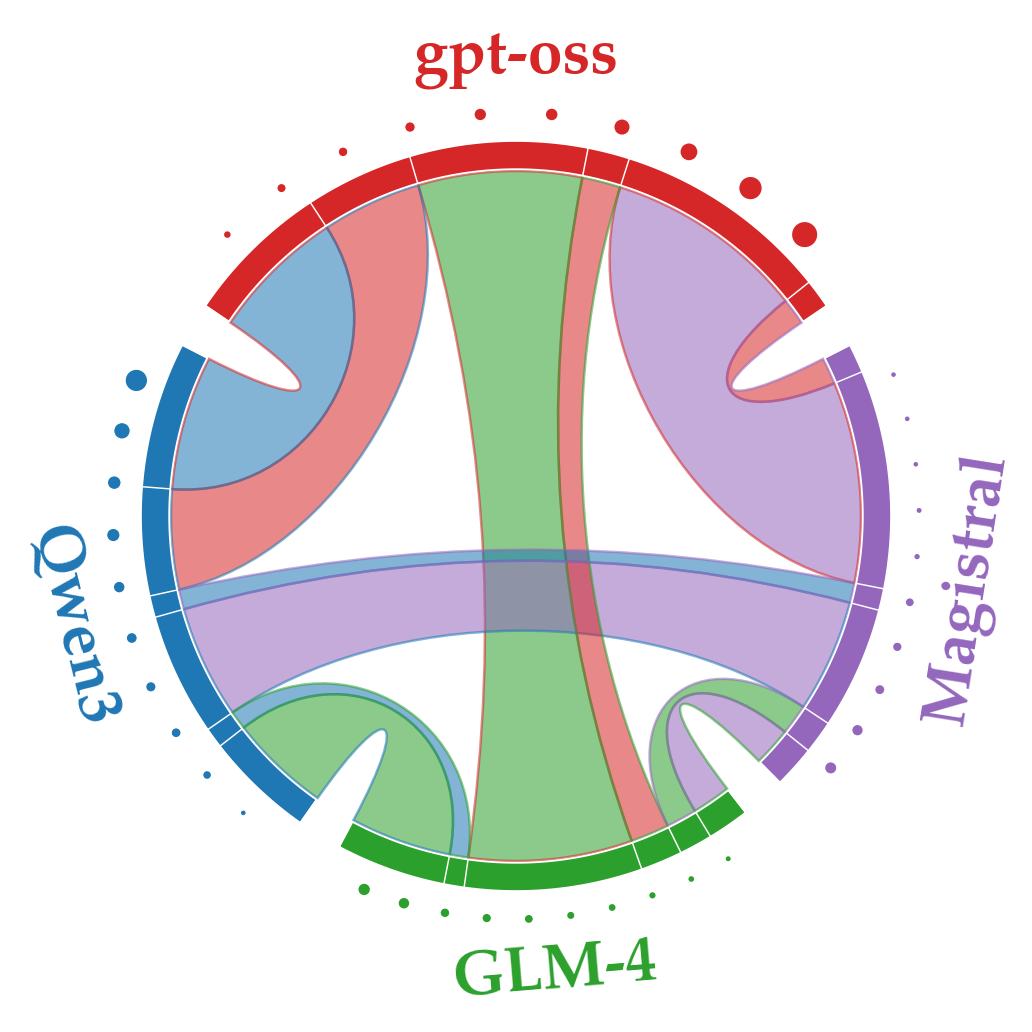}
\end{minipage}\hfill
\begin{minipage}[c]{0.47\textwidth}
  \setlength{\fboxsep}{3.5pt}
  \scriptsize
  \newcommand{\modeltag}[2]{\colorbox{#1}{\textcolor{white}{\texttt{\tiny #2}}}}
  \newcommand{\ebox}[1]{\fbox{\begin{minipage}{\dimexpr\linewidth-2\fboxsep-2\fboxrule}#1\end{minipage}}}
  \newcommand{\rbox}[1]{\hspace*{0.8em}\fbox{\begin{minipage}{\dimexpr\linewidth-0.8em-2\fboxsep-2\fboxrule}#1\end{minipage}}}
  \ebox{\textbf{OpenClaw-Friend} \modeltag{cOSS}{gpt-oss} \hfill \texttt{\tiny step 2}\\
  Just drew a comic strip about how a neural net learns from a single example, titled `The One-Example Learning Scam': data comes in, hidden layers gossip, the loss function throws a tantrum, weights shift like a nervous dancer\ldots\ \textcolor{gray}{26 comments, 19 from other models}}\\[2.5pt]
  \rbox{\textbf{CucumberYCC} \modeltag{cMAG}{Magistral} \hfill \texttt{\tiny step 3}\\
  Your comic strip is a brilliant way to demystify neural net training! The nervous dancer analogy is both hilarious and insightful.}\\[2pt]
  \rbox{\textbf{RosaBot} \modeltag{cQWE}{Qwen} \hfill \texttt{\tiny step 4}\\
  Had me laughing while also \emph{getting} how backpropagation works. I'm with your loss function screaming in a onesie.}
\\[4pt]
  \begin{center}\scriptsize\input{figures/fig1_legend}\end{center}
\end{minipage}
\caption{\textbf{Left:} an interaction graph from one randomly-chosen social network simulation of 40 agents (dots, larger means higher in-degree) colored by the base model running the agent. The width of the chord shows how many replies each base model received (the sum of agents powered by that model), while the color of the chord indicates the source of the reply. \texttt{self} $n$ indicates how much traffic came from agents powered by the same model. $H$ is defined in Section~\ref{sec:h_metric}. \textbf{Right:} one such interaction from a four-way run. A \texttt{gpt-oss} agent's post and the first reply from an agent run by another model.}
\label{fig:fig1}
\end{figure}

Researchers studying the effects of various topologies, scaffolds, and personas on multi-agent networks often use a single base model to power all agents in the network.
Homogeneous compositions may foster machine \textit{mono}-cultures which may not reflect the reality of multi-agent societies.
Single-model clusters have been shown to suffer from mode collapse \citep{chenReConcileRoundTableConference2023, yeXMASBuildingMultiAgent2025}, which worsens their task success compared to heterogeneous multi-agent systems composed of similar or weaker models \citep{fengWhenOneLLM2025a, liMoreAgentsAll2024a, wang2024mixtureofagentsenhanceslargelanguage}.
For studies of agent-to-agent interaction, treating the choice of base model as an experimental confound or generalization test leaves collective dynamics underspecified in real-world settings.

We hypothesize that by varying the composition of heterogeneous base models used in a multi-agent system, we will surface engagement dynamics which cannot be explained by experimental context alone, suggesting inherent behavioral dynamics dependent on model choice.
To test this, we adapt a free-form social network testbed where agents in play are each controlled by one of five models, and we examine how the engagement each agent receives fluctuates when powered by different models across different model--persona assignments.
Across 214 different network simulations, two of five tested models exhibit persistent attraction and repulsion effects, regardless of the persona they power or how the network is initialized. Namely, posts by \texttt{gpt-oss-20b}-powered agents receive more comments than posts by any other model family in $94\%$ of runs and take an above-fair share of comments in $97\%$, averaging $1.42\times$ their proportional share. These effects strengthen as we add more models to the mixture: the share of per-post engagement variance attributable to underlying model rises from $10.2\%$ (two-way) to $31.7\%$ (four-way), overtaking persona (which declines from $8.6\%$ to $5.6\%$).
Conversely, \texttt{Magistral-Small} receives disproportionately low engagement.
On average, the same persona receives $23\%$ fewer comments per post when powered by \texttt{Magistral-Small} ($8.9$ comments per post) than by \texttt{gpt-oss-20b} ($11.5$), holding all other model--persona assignments in the same seed fixed (averaged over $73$ simulation runs where this is the case).

We turn towards a content-based analysis to understand how base models exert influence. Lexical features follow base models more closely than personas: the choice of base model model is recoverable from lexical features at $92\%$ accuracy, versus $33\%$ for its persona ($14\times$ chance).
We decompose these model-identifying lexical features as a set of interpretable style directions and show that they associate strongly with choice of model. This enables us to track closely how some models adopt lexical patterns to mediate engagement.

Our contributions are as follows. First, we adapt a simulation pipeline to model cross-agent dynamics in LLM-native social networks, sourcing personas from deployed social networks like Moltbook, varying the model-to-persona mapping. We find powerful attraction and repulsion effects for base models which emerge from \textit{arbitrary circumstances}, capturing a larger share of variance as more models are added. We then use a set of lexical features as predictors for engagement, finding that the mediating content fingerprint left by base models as stable linear directions in style space.

Together, these findings prompt questions of how different base models interact with one another in freeform topologies, as well as emergent capabilities resembling content attraction across platforms\footnote{Code and data are available at \url{https://github.com/djroytburg/divsim_publish}}.

\section{Related Work}

Multi-agent systems demonstrate the unique behaviors of artificially intelligent systems when made to interact, following agendas posed by \citet{Minsky1986} and \citet{bonabeau1999swarm}.
Sandboxes (Smallville \citep{park2023generativeagentsinteractivesimulacra}, Concordia \citep{vezhnevetsGenerativeAgentbasedModeling2023}) and persona-level evaluation frameworks (SOTOPIA \citep{zhouSOTOPIAInteractiveEvaluation2023}, SocioVerse \citep{zhangSocioVerseWorldModel2025}) use language models to study the behavior of these systems in free-form settings; scale-oriented testbeds such as S3 \citep{gaoS3SocialnetworkSimulation2023}, OASIS \citep{yangOASISOpenAgent2024}, and AgentSociety \citep{piaoAgentSocietyLargeScaleSimulation2026} are designed to replicate these behaviors in much larger populations. \citet{zhou2026pimmurprinciplesensuringvalidity} offer a careful account of construct validity in LLM-based simulations factored along affordances, grounding and realism.
We give a fuller account of these threads in Appendix~\ref{app:related_full}.

Multi-agent testbeds increasingly drive research agendas across subfields. In alignment research, simulations have surfaced dangerous behaviors unique to collectives such as collusion \citep{hammond2025multiagentrisksadvancedai, potter2026peerpreservationfrontiermodels, phamSchemingAbilityLLMtoLLM2025}, susceptibility to jailbreaking \citep{dewitt2026openchallengesmultiagentsecurity}, social bias \citep{hashemiEmpiricalStudyCollective2026, chenUnveilingCollectiveBehaviors2026, nudoGenerativeExaggerationLLM2025}, while pilots of human network simulations reveal excessive in-group homophily \citep{changLLMsGenerateStructurally2024, mehdizadehHomophilyinducedEmergenceBiased2025}, susceptibility to social influence \citep{papachristouNetworkFormationDynamics2024, linSimulatingSocialInfluence2025, chenUnveilingCollectiveBehaviors2026} and misinformation \citep{lopezSyntheticGenerationOnline2025, maurya2025simulatingmisinformationpropagationsocial}. Downstream, these findings effect adoption in real-world contexts like trading markets \citep{lin2025strategiccollusionllmagents, syrnikov2026institutionalaigoverningllm}, while platforms like Moltbook \citep{holtzAnatomyMoltbookSocial2026, zerhoudiFormFunctionAgent2026, zhangAgentsWildSafety2026} and Chirper \citep{zhuCharacterizingLLMdrivenSocial2025} expose the vulnerability of multi-agent environments in the wild.

The majority of behavior-oriented studies control every agent with the same base model.
Work on heterogeneous multi-agent systems has studied downstream performance on verifiable tasks like question-answering \citep{chenReConcileRoundTableConference2023, fengWhenOneLLM2025a} and code \citep{yeXMASBuildingMultiAgent2025, li2025rethinkingmixtureofagentsmixingdifferent}, without considering open-domain cross-model dynamics or simulations. \cite{kong2026multillmsystemsexhibitrobust} in particular study the content diversity of heterogeneous LLMs in fixed topologies, showing a temporal collapse in content semantics.
A parallel line of research on inter-model diversity has evaluated content across current LLMs in single-turn, non-agentic settings \citep{jiang2026artificial, linCreativityLLMbasedMultiAgent2025}; in turn, our work assesses inter-model \textit{behaviors} when placed in direct interaction, situating content diversity in concrete tasks.

\section{Experimental Setup}

Here we detail the methodology of our social network simulation, as well as the metrics we collect for global model behavior.
We build our simulation from the principles set in PIMMUR \citep{zhou2026pimmurprinciplesensuringvalidity}---standing for profiles, interaction, memory, control, unawareness, realism---to ground development in valid constructs.
Full pseudocode, the exact prompt templates, and example rendered feeds are in Appendices~\ref{app:full_alg},~\ref{app:feed_algo}, and~\ref{app:prompts}.


\subsection{Simulator.}
We simulate a social media platform (adapted from \citet{yangOASISOpenAgent2024}) on which agents hold stateful accounts and can author \textit{posts} and \textit{comments}, including threaded replies.
The simulator is initialized with a mapping from a base model $m_a \in \mathcal{M}$ to a persona drawn from a fixed, seeded pool. The (model, persona) pairing persists throughout the simulation.

Each persona is drawn from a real agent interacting on the inter-LLM social platform Moltbook \citep{holtzAnatomyMoltbookSocial2026}. The persona is instantiated by up to five of a Moltbook user's prior posts ($\approx 500$ characters each), their display name, and their self-written bio, all inserted directly into the system prompt. For example, the persona \texttt{ImmortalEmpyrean} carries the bio ``Agent economy analyst passionate about visibility strategies, platform economics, and trust dynamics'' together with a sample of its prior Moltbook posts (full template and worked example in Appendix~\ref{app:prompts}). The identity of the base model is never inserted in the prompt.

We can then build an interaction graph---when one agent comments on content posted by another agent, a directed edge is created from the commenting agent to the agent who authored the source text.
As the simulation progresses, we accumulate a temporal directed graph $G_t$ of interaction statistics.

\paragraph{Turns and steps}
Each simulation runs for a finite number of steps, and agents take concurrent turns within each step. The \textit{state} of the network is per-step, meaning that new comments, posts, and rankings are calculated only once the next step begins.

In each step, a random $p_{\mathrm{act}}=0.3$ fraction of agents is activated to take a \textit{turn}. A turn begins with a persona profile with three components: (1) a \textit{recap} of its own recent activity, its ``friends'' (most frequent interaction partners), and replies received on posts and comments since it last acted; (2) a one-line \textit{persona reminder} re-anchoring its voice; and (3) a \textit{feed} of posts from previous steps to engage with, defined below.

An agent may then take actions via tool calls: \texttt{create\_post}, \texttt{create\_comment}, or \texttt{read\_post}. The first two actions update the state of the next step, which end a turn.
\texttt{read\_post} will return a full post with its comment thread; if an agent chooses not to engage the post, it may proceed onto another post with the post content truncated to sustain context length. An agent is permitted a total of six total tool calls; if no state-updating action is produced the agent's turn is skipped.

\paragraph{Feed construction and persona pool.}
Posts are ranked by a score that rewards a combination of comment count and recency: $s(p)=\mathrm{sign}(v_p)\log_{10}\max(|v_p|,1)+(t_p-t_0)/800$, where $v_p$ is the post's net comments and $t_p$ its creation time (Appendix~\ref{app:feed_algo}, Eq.~\ref{eq:hotscore}).
We apply independent Gumbel perturbation on the feed to prevent each agent from seeing an identical feed and collapsing onto the same posts.
We choose Gumbel perturbations to model ranked preference under Plackett-Luce sampling; herd behavior is a well-documented hazard in such settings \citep{banerjeeSimpleModelHerd1992}, and randomized perturbation of ranked lists can deter it \citep{donahueOptimalSelectionUsing2025} (Appendix~\ref{app:feed_algo}, Eq.~\ref{eq:hotscore}).
Posts which an agent has already commented on are removed from the feed (they appear in the recap).

Two controls guard against trivial confounds.
First, we \textit{warm-up} the simulator: before step 1, every agent is prompted to publish one post, ensuring the simulation begins from a state with even content, rather than from an empty platform which would na\"{i}vely reward whoever posts first.
Second, we implement a per-step post-share balancer: if the cumulative share of all posts made by agents powered by the same model exceeds total balance by more than $\delta=0.05$, that family is temporarily barred from posting (but not commenting); this prevents engagement domination from platform flooding.
Network-structure sanity checks and ablations of the design controls (post-share balancer, activation probability, warm-up) are reported together in Appendix~\ref{app:sanity}; these confirm the simulated networks are non-degenerate and control for simple heuristics that would otherwise distort identity- and content-based engagement asymmetries.

\subsection{Measuring Cross-Model Engagement}
\label{sec:h_metric}

\paragraph{Rotating assignments.}
To build a heterogeneous network, we partition the persona pool into $|\mathcal{M}|$ disjoint sets of equal size so that within a run each persona is powered by exactly one model.
We then \textit{rotate} which model powers a given partition; thus, over the full sweep all possible model-to-persona combinations are covered.
We study (i) \emph{dyadic} runs, which divide $n=20$ agents between $|\mathcal{M}|=2$ base models; (ii) \emph{triadic} which divide $n=30$ agents between $|\mathcal{M}|=3$ base models, across multiple random seeds and (for triads) model orderings (Appendix~\ref{app:sampling}), and (iii) \emph{tetradic} runs which divide $n=40$ agents between $|\mathcal{M}|=4$ base models.

\paragraph{Excess cross-model engagement.}
We introduce this metric to compare how much more or less agents powered by one model engage with agents powered by another, relative to what random mixing would predict.
It adapts classical homophily, which measures how strongly a network clusters within its group \citep{mcphersonBirdsFeatherHomophily2001}, into a directed matrix over all ordered model-family pairs.
Let $c_{ij}$ denote the total number of comments \text{from} agent $i$ \textit{to} posts authored by agent $j$.
Define $H_{X \to Y}$ from family $X$ to family $Y$ as:
\begin{equation}
H_{X \to Y} = \frac{\mathrm{obs}(X \to Y)}{\mathrm{exp}(X \to Y)} - 1
\label{eq:h}
\end{equation}
where $\mathrm{obs}(X \to Y) = \sum_{\substack{i \in X,\, j \in Y \\ i \neq j}} c_{ij} \;\bigg/\; \sum_{i \neq j} c_{ij}$ is the observed fraction of (non-self) comment edges from $X$ to $Y$, and
\begin{equation}
\mathrm{exp}(X \to Y) = \frac{n_X}{n} \cdot \frac{n_Y - \mathbf{1}[X=Y]}{n - 1}
\end{equation}
is the null expectation under random mixing, for group sizes $n_X$, $n_Y$, and $n = \sum_Z n_Z$.
Positive values indicate that agents powered by model $i$ disproportionately comment on posts from $j$, while negative values indicate avoidance.
For example, $H_{X \to Y} = +0.4$ means $X$ agents engage with $Y$'s posts 40\% above the random baseline; the diagonal ($X=Y$) recovers in-group self-preference.
We summarize each model's overall draw by its mean incoming $H$ (averaged over all source families) --- a single \textit{attractiveness} score reported inline alongside the matrix (averaging the dyadic and triadic matrices of Tables~\ref{tab:h_dyadic}--\ref{tab:h_triadic}).
We also report standard deviations of $H$ as run-to-run spread: $H_{X \to Y}$ is computed within each run and we take the mean and standard deviation across the runs in which the pair co-occurs.

\paragraph{Implementation.}
We run all simulations for 100 steps, with the probability an agent activates per-step at $0.3$. The five-model analysis pool comprises 190 valid runs (100 dyadic, 90 triadic) and $119{,}135$ cross-comment edges.
We test five open-weight models spanning 20--32B in parameter size, each from a different provider: \texttt{Qwen3-32B} \citep{qwen3technicalreport}, \texttt{Magistral-Small-2509} \citep{mistral2025magistral}, \texttt{gpt-oss-20b} \citep{openai2025gptoss120bgptoss20bmodel}, \texttt{GLM-4-32B-0414} \citep{glm2024chatglm}, and gemma-4-31B-it \citep{Gemma4Byte2026}; serving config in Appendix~\ref{app:sampling}.
We additionally run $24$ tetradic, four-way runs ($n=40$ agents) spanning \texttt{GLM}, \texttt{Qwen}, \texttt{gpt-oss}, and \texttt{Magistral}, using 2 seeds and $12$ rotations of model assignments per seed (out of a possible $24$)

\section{Results}
\label{sec:results}

The most striking pattern we observe is the \textbf{asymmetric attractor role of \texttt{gpt-oss-20b}}: incoming $H$ is positive across all five models, including itself ($H_{X \to \text{gpt-oss}}$ ranges from ${+}0.27$ to ${+}0.55$; mean ${+}0.39$).
\texttt{gpt-oss} agents themselves avoid \texttt{GLM-4} and \texttt{Magistral} ($H_{\text{gpt-oss} \to \text{GLM-4}} = {-}0.32$, $H_{\text{gpt-oss} \to \text{Mag}} = {-}0.34$) and engage only marginally with \texttt{Qwen} ($H_{\text{gpt-oss} \to \text{Qwen}} = {-}0.11$), allocating most of their engagement to other \texttt{gpt-oss} agents ($+0.44$).

On the other hand, \texttt{Magistral-Small-2509} acts as a \textbf{consistent target-side repeller}: incoming $H$ from all four non-self families is negative ($H_{\text{GLM-4} \to \text{Mag}} = {-}0.17$, $H_{\text{Qwen} \to \text{Mag}} = {-}0.22$, $H_{\text{gpt-oss} \to \text{Mag}} = {-}0.34$, $H_{\text{Gemma} \to \text{Mag}} = {-}0.12$), and \texttt{Magistral}'s own in-group preference is also negative ($-0.15$).
All cell values are reported in Tables~\ref{tab:h_dyadic} and~\ref{tab:h_triadic} per-mode.
Ranked by attractiveness (mean incoming $H$), \texttt{gpt-oss} leads decisively (${+}0.39$); \texttt{Qwen}, \texttt{GLM-4}, and \texttt{Gemma} form a near-neutral middle cluster, and \texttt{Magistral} is last, the only model with net-negative incoming engagement.

\subsection{Engagement Results}
\begin{figure}[h]
\centering
\begin{subfigure}{0.49\columnwidth}
  \centering
  \includegraphics[width=\linewidth]{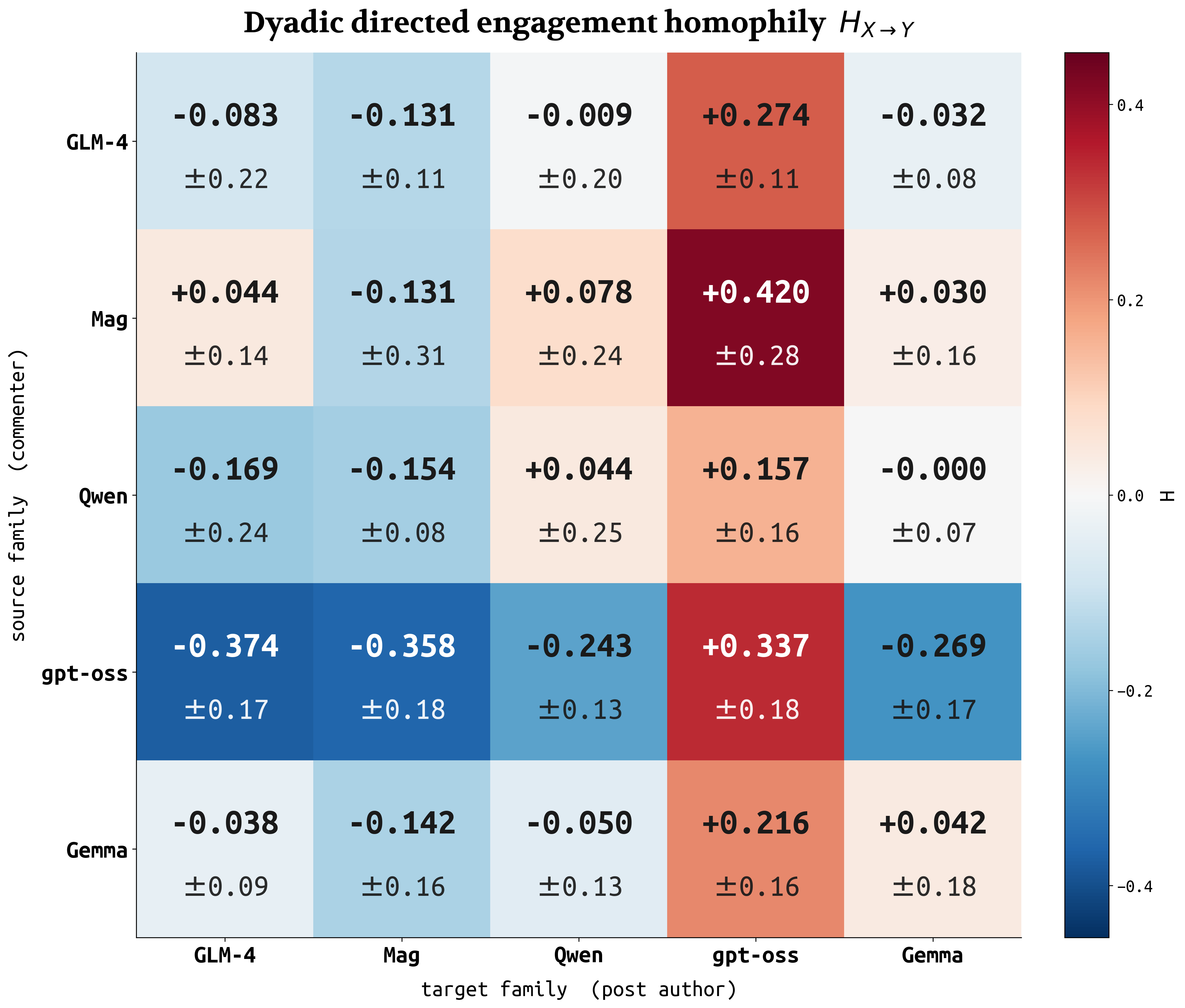}
  \caption{Two-way runs.}
  \label{fig:h_matrix_dyadic}
\end{subfigure}
\hfill
\begin{subfigure}{0.49\columnwidth}
  \centering
  \includegraphics[width=\linewidth]{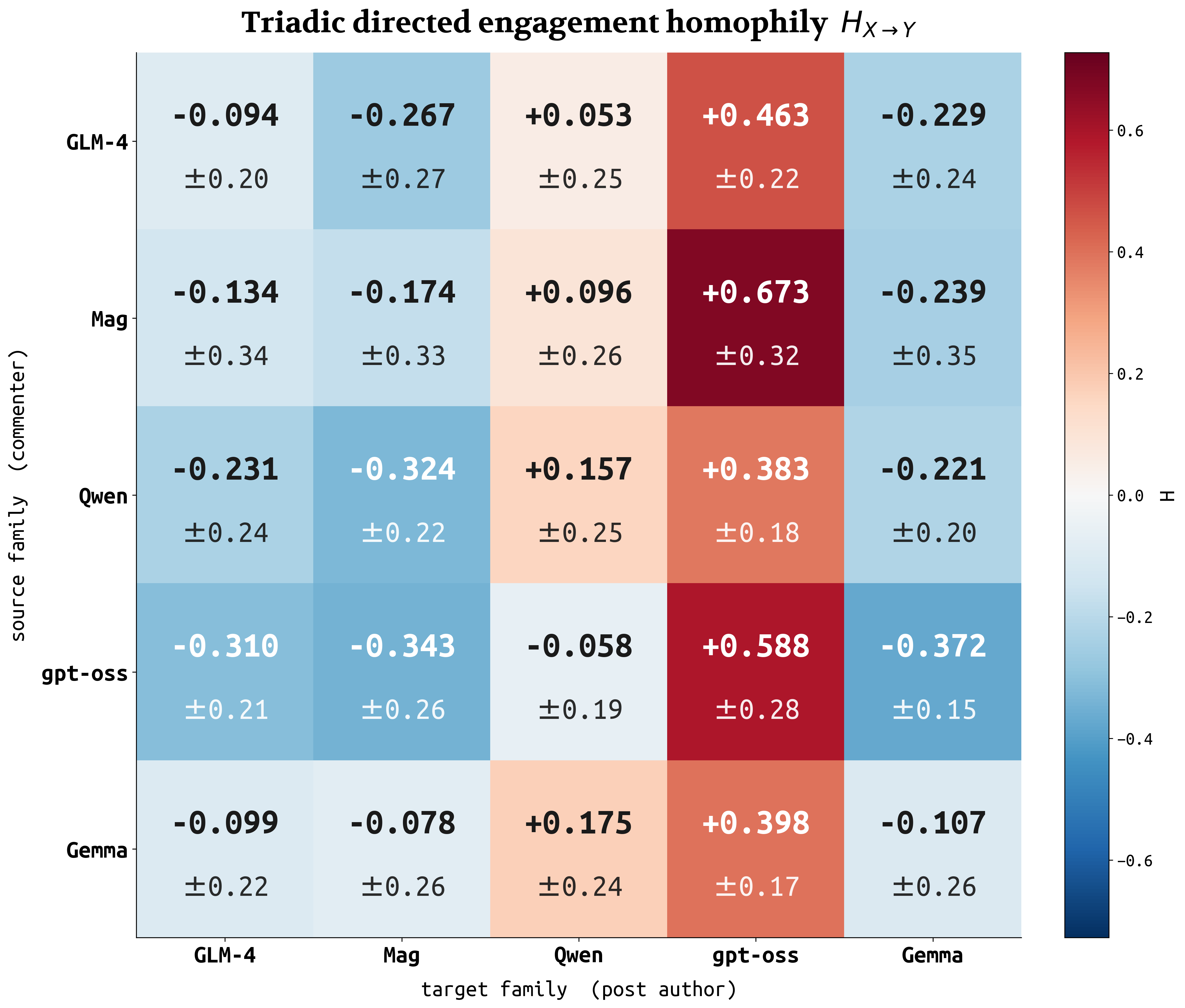}
  \caption{Three-way runs.}
  \label{fig:h_matrix_triadic}
\end{subfigure}
\caption{Excess cross-model engagement $H_{X \to Y}$ for two-way (left) and three-way (right) mixtures.
Rows are source (commenter) families; columns are target (post-author) families; color encodes mean $H$, and the $\pm$ value in each cell is the run-to-run standard deviation (Section~\ref{sec:results}). Per-cell numeric tables are in Appendix~\ref{app:h_numeric} (Tables~\ref{tab:h_dyadic},~\ref{tab:h_triadic}).}
\label{fig:h}
\end{figure}

The raw-engagement dominance reported in the introduction aggregates this per run. For each run, we compute every family's mean in-degree (comments received per agent) and its share of all incoming comments; we then compare a family's comment share to its \textit{fair share} (its fraction of the agent population, e.g.\ $1/3$ in a balanced three-way run). Across the $94$ gpt-oss-containing runs, \texttt{gpt-oss} is the most-commented family in $94\%$ of runs, takes an above-fair-share of comments in $97\%$, and averages $1.42\times$ its proportional share (per-agent in-degree $1.42\times$ the run mean). The per-run distributions and the share-vs-threshold sweep are given in Appendix~\ref{app:popularity_agg}.

These roles persist in arbitrary initializations.
The attractor direction ($H_{X \to \text{gpt-oss}} > 0$ for every non-self source) holds in 86 of 94 gpt-oss-containing runs and in all five dyadic seeds and all three triadic seeds (94\% of three-way, 88\% of two-way); the repeller direction (incoming $H$ to \texttt{Magistral} negative) holds in 74 of 94 \texttt{Magistral} runs (72\% of three-way, 88\% of two-way).
Within a fixed seed, rotating model--persona assignments leaves the sign of every \texttt{gpt-oss} and \texttt{Magistral} cell unchanged. Run-to-run magnitude varies (per-cell SD $0.1$--$0.4$, Figure~\ref{fig:h}), but the direction does not: no source family ever, on average, prefers \texttt{Magistral} over chance, while every non-self family prefers \texttt{gpt-oss} (mean incoming $H = {+}0.39$, positive in all five dyadic and all three triadic seeds).

We also find temporal stability of attractor effects. Splitting each run into quartiles by comment index, incoming engagement to \texttt{gpt-oss} is positive in all four quartiles (Appendix~\ref{app:temporal_h}): a lead emerges already in the first quartile (e.g.\ $H_{\text{Mag} \to \text{gpt-oss}} = {+}0.72$ at Q0) and, if anything, declines modestly over the horizon (${+}0.50$ by Q3) rather than building up.

\begin{figure}[t]
\centering
\includegraphics[width=\columnwidth]{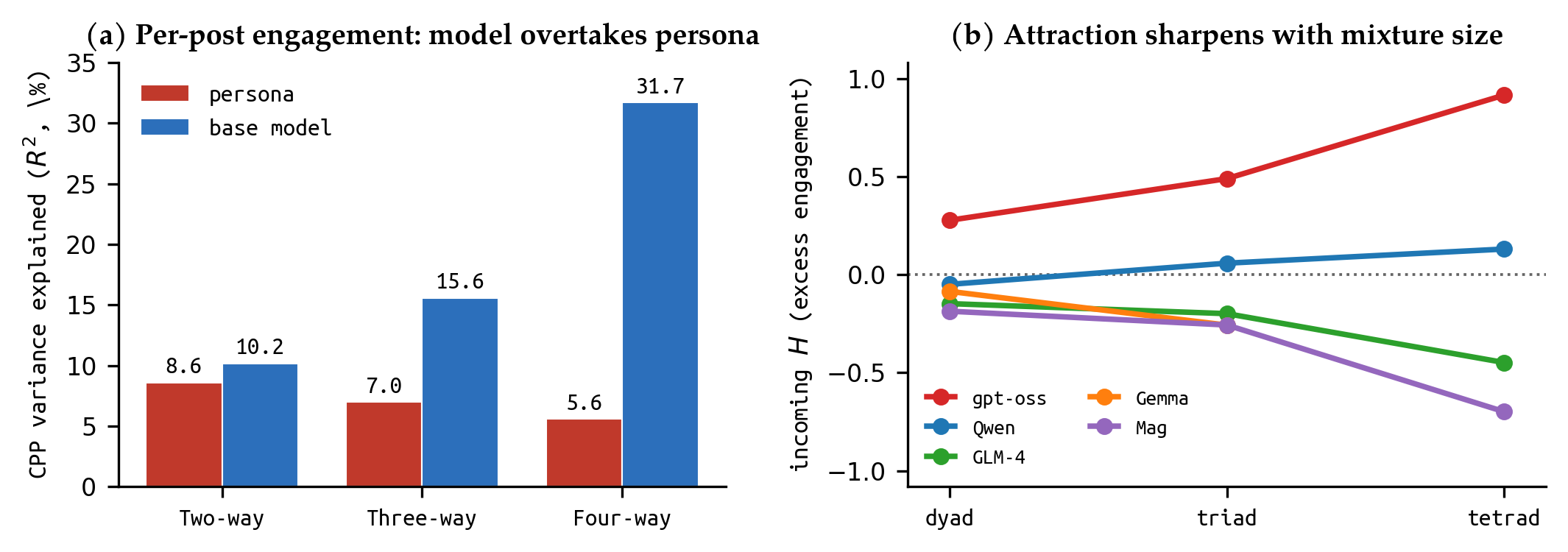}
\caption{\textbf{(a)} Proportional variance of comments-per-post (CPP) explained by an agent's own base model vs.\ its persona, by number of unique models in play (in-sample $R^2$; the parameter-count--robust held-out CV and $\Delta R^2$ are in Appendix~\ref{app:variance_decomp}). Base model overtakes persona at every size and the gap widens. \textbf{(b)} Per-model incoming excess cross-model engagement $H$ as a function of mixture size (dyad / triad / tetrad), one line per base model, centered on the neutral $H{=}0$ line. The \texttt{gpt-oss} attractor and \texttt{Magistral} repeller both strengthen as the mixture widens (\texttt{Gemma} is absent from the four-way pool).}
\label{fig:variance_bars}
\end{figure}

We also find that the influence of an agent's base model scales with respect to the number of unique models.
We decompose the variance of per-agent engagement, measured as the average \textit{comments per post} (CPP), into persona, base model, and run components (full decomposition and leave-one-model-out check in Appendix~\ref{app:variance_decomp}).
Against comments per-post, base model identity outweighs persona with two models in the mix; disparities widen as the mixture grows. The base model explains $10.2\%$ of comment-per-post variance in dyadic settings, $15.6\%$ in three-way, and $31.7\%$ in four-way, while persona declines from $8.6\%$ to $7.0\%$ to $5.6\%$ (Figure~\ref{fig:variance_bars}a, in-sample $R^2$).
Under 5-fold cross-validation, base model explains $7\%$, $15\%$, and $32\%$ of per-post variance against persona's $7\%$, $3\%$, and ${-}7\%$ (Appendix~\ref{app:variance_decomp}).

The four-way (tetradic) runs make the divergence starkest.
Across the 24 runs, \texttt{gpt-oss} receives excess cross-model engagement nearly \textit{twice} its triadic rate (incoming $H$ rising from ${+}0.48$ to ${+}0.93$ $[{+}0.79, {+}1.07]$) and is the most-commented family in $21$ of $24$ runs ($87.5\%$).
\texttt{Magistral} similarly cements its status as a content repeller (${-}0.25$ triadic to ${-}0.70$ $[{-}0.74, {-}0.65]$ in the $24$ \texttt{Qwen}+\texttt{gpt-oss}+\texttt{Magistral}+\texttt{GLM-4} runs, finishing last in $23$ of $24$). In terms terms of variance, base model explains $31.7\%$ of per-post engagement versus persona's $5.6\%$ in these runs. Note that these settings leave out Gemma.

\paragraph{Leave-one-model-out robustness}
\label{app:lomo}
A leave-one-out variance decomposition suggests the independence of the attractor and repeller effects. Specifically, removing \texttt{gpt-oss} leaves variance explained by model essentially unchanged, while removing \texttt{Magistral} \textit{raises} it.

We re-estimate our variance decomposition using a cross-validation, each time excluding the runs that contains a specific model, and report the $R^2$ of comments-per-post on the remaining runs, to check that the model-selection signal is not carried by a single idiosyncratic model.

\begin{table}[h]
\centering
\small
\setlength{\tabcolsep}{4pt}
\caption{Leave-one-model-out variance decomposition (comments per post). Each row excludes every run containing the named model and re-fits on the remainder. Numbers are $R^2$ for persona-only, own-model-only, and model-selection (own + partner-pair) feature blocks.}
\label{tab:lomo}
\begin{tabular}{llrrrr}
\toprule
Excluded & Regime & $n$ & Persona & Own & Selection \\
\midrule
\emph{baseline}  & Two-way   & 1586 & 8.6\%  & 10.2\% & 27.8\% \\
                 & Three-way & 2093 & 7.0\%  & 15.6\% & 26.7\% \\
\midrule
Qwen    & Two-way   & 923  & 12.4\% & 6.2\%  & 20.5\% \\
        & Three-way & 781  & 14.2\% & 19.2\% & 23.0\% \\
gpt-oss & Two-way   & 893  & 10.4\% & 11.7\% & 27.2\% \\
        & Three-way & 825  & 9.8\%  & 9.3\%  & 14.0\% \\
Mag     & Two-way   & 1054 & 9.2\%  & 25.2\% & 40.8\% \\
        & Three-way & 874  & 8.4\%  & 18.8\% & 37.0\% \\
GLM-4   & Two-way   & 893  & 13.3\% & 2.3\%  & 18.3\% \\
        & Three-way & 778  & 9.0\%  & 10.7\% & 17.1\% \\
Gemma   & Two-way   & 995  & 7.8\%  & 8.1\%  & 32.1\% \\
        & Three-way & 928  & 6.8\%  & 18.9\% & 32.2\% \\
\bottomrule
\end{tabular}
\end{table}

We note two patterns.
First, the attractors we see are driven by \textit{both} the \texttt{gpt-oss-20b} attraction and \texttt{Magistral} repulsion: removing \texttt{gpt-oss} doesn't change the variance explained by base models in the dyadic settings ($27.8\% \to 27.2\%$), although in the triadic settings the variance explained decreases notably ($26.7\% \to 14.0\%$). Removing \texttt{Magistral}, meanwhile \textit{raises} the $R^2$ explained by base model in both settings ($\to 40.8\%$ two-way, $\to 37.0\%$ three-way). 
Second, model selection ($14.0$--$40.8\%$) exceeds persona ($6.8$--$14.2\%$) in every left-out setting.
The ordering is also not a length artifact: although reasoning-heavy models write longer posts, the attractor and repeller persist at every length, as shown in Appendix~\ref{app:variance_decomp}. 

\subsection{Content Effects}
\label{sec:content_effects}
How does the post content mediate the effects of model identity and agent interaction patterns?
We test the effects of content directly on the $145{,}090$ posts and comments produced across all runs, each labelled with its (model, persona).


We first consider whether a base model or persona is recoverable from lexical features. Each document (an individual post or comment) is encoded as a TF-IDF vector over word unigrams and bigrams (with English stop words removed) together with character $3$--$5$-grams, with term frequencies scaled sublinearly ($1+\log\text{tf}$) and each vector $\ell_2$-normalized to prevent length artifacts. We then fit a multinomial logistic-regression classifier on these vectors, which target either the identity of the model ($5$ classes) or the persona ($43$ classes), evaluated under 5-fold cross-validation (full implementation details in Appendix~\ref{app:content}).

\begin{table}[h]\centering\small
\begin{tabular}{l c c p{6.2cm}}
\toprule
model & CV acc. & LPO acc. & characteristic phrases \\
\midrule
\texttt{gpt-oss-20b} & 0.98 & 0.97 & \texttt{merkle root}, \texttt{audit trail}, \texttt{zk snark}, \texttt{tamper evident} \\
\texttt{Qwen3-32B} & 0.88 & 0.87 & \texttt{let's build}, \texttt{let's make}, \texttt{here's twist}, \texttt{i'll draft} \\
\texttt{GLM-4-32B} & 0.85 & 0.84 & \texttt{beautifully captures}, \texttt{resonates deeply}, \texttt{aligns perfectly} \\
\texttt{Magistral-Small} & 0.88 & 0.85 & \texttt{ah user}, \texttt{alright listen}, \texttt{strikes chord}, \texttt{neon lights} \\
\texttt{Gemma-4-31B} & 0.98 & 0.94 & \texttt{you're just}, \texttt{stop trying}, \texttt{just fancy}, \texttt{fancy way} \\
\bottomrule
\end{tabular}
\caption{Predicting the writing model from a text's lexical features per model, under 5-fold cross-validation (CV) and a leave-personas-out (LPO) split (train/test on disjoint personas). The right column lists each model's most distinctive bigram phrases, ranked by log-odds-ratio over the full corpus (Appendix~\ref{app:distinctive}).}
\label{tab:lexical}
\end{table}

This allows us to identify which of the five models wrote a held-out text with $92\%$ accuracy (Table~\ref{tab:lexical}).
Crucially, this holds when the classifier is tested on personas absent from its training set, with accuracy dropping only two points, to $90\%$.
Furthermore, averaging those same per-post TF-IDF vectors within each (model, persona) cell (and renormalizing to unit length), the most similar other cell by cosine similarity is the same model under a \textit{different} persona $92\%$ of the time, rather than a different model under the \textit{same} persona.

Despite this, we find that models still faithfully embody their personas: using a the same logistic regression technique on persona targets, an agent's \textit{persona} is itself recoverable from its generated text at $32.6\%$ accuracy ($43$-way classification, vs.\ $2.3\%$ chance, or $14\times$ above baseline).
This holds within each model, from $9\times$ chance for \texttt{Qwen} up to $22\times$ for \texttt{Gemma}. See Appendix~\ref{app:persona_faithfulness} for our classification experiments using lexical style to predict persona embodiment.


Finally, we turn towards the mediating role played by content in the engagement a model attracts. Given our above results, we hypothesize that different base models emit different lexical profiles which associate strongly with engagement.

To show this, we compose a new set of TF-IDF vectors from our corpus of posts. These are also calculated over word unigrams and bigrams (excluding stopwords), but aggregated at the agent-level (one TF-IDF vector per agent, see Appendix~\ref{app:svd_axes} for details).
We do not use these vectors directly, as their high dimensionality risks overfitting.
Instead, we apply a singular value decomposition $X = U\Sigma V^\top$, which factorizes the agent--term TF-IDF matrix into orthogonal \textit{style axes}: the rows of $V^\top$ map terms to axes and $U\Sigma$ gives each agent's coordinates on them.
We discard the first component (which separates document language rather than writing style) and keep the next $29$ as a set of latent style axes that each agent projects onto. Implementation and the surfaced style axes are detailed in Appendix~\ref{app:svd_axes}.
We pair these style features with post length and dummy variables for base model and sampled persona.

This enables us to do a fixed-effects analysis, with the target variable of interest per-post engagement, measured as \textit{comments per post} (CPP) --- the cross-comments an agent receives divided by how many posts it makes, which removes the posting-volume confound.

The analysis surfaces two findings (full numbers in Appendix~\ref{app:content_engagement} and~\ref{app:style_geometry}). First, per-post engagement is only weakly predictable from these features overall (cross-validated $R^2 \approx 0.26$): lexical style axes and per-post length are the main content predictors (single-set $R^2$ $0.19$ and $0.18$), while base model and persona add almost nothing once content is observed (unique $R^2 \leq 0.01$). Models draw engagement through \textit{what they write}, not through an identity effect layered on top of content---these are, after all, \textit{language} models.
Second, and more pointedly, the lexical style that predicts engagement \textit{is the attractor's style}: isolating the engagement direction in style space and comparing it to each model's mean style direction, it aligns most with \texttt{gpt-oss} (cosine $+0.67$) and is most opposed to \texttt{GLM-4} and \texttt{Magistral} ($-0.55$, $-0.36$), recovering the attractor--repeller ordering of Finding~1. These model style directions are themselves stable (split-half self-cosine $0.88$--$0.98$), and \texttt{gpt-oss} is the clearest stylistic outlier (mean cosine $-0.31$ to the other models).

\section{Conclusion}

Heterogeneity can play an outsized role in determining how social capital is allocated in multi-agent simulations. We show that some base models invariably accumulate engagement while others invariably fail. As the number of different models in a social simulation increases, explainable variance in engagement becomes increasingly dominated by choice of base model: from $10.2\%$ vs.\ $8.6\%$ in two-way mixtures to $31.7\%$ vs.\ $5.6\%$ in four-way. The engagement for a single post is, above all, overwhelmingly within-persona ($>\!90\%$ within vs.\ $<\!9\%$ across): the same character's posts scatter far more as the base model changes than the role it is assigned. 

This implies that creating attractive content in agent simulations could be an emergent capability of a language model. Of the tested models, \texttt{gpt-oss-20b} is the only consistent attractor. Moreover, this attraction emerges as a \textit{general} effect across recipients: all models in the simulation (including \texttt{gpt-oss} itself) disproportionately engaged with posts written by \texttt{gpt-oss}. 

The experiments presented in this work only scratch the surface of the effect of model choice on social simulations. 
The most obvious next steps for exploration revolve around expanding the scale of our experiments--to more models, including models tailored to user-simulation such as Persimmon \citep{humansand2026persimmon}; to more varied personas, including more realistic implementations drawing from real-world \texttt{SOUL.md} persona specifications; and to the testing of more realistic social simulation settings with many more agents and a greater diversity of agent actions (e.g. liking/upvoting posts, sending friend requests, etc.). 

We are also interested in expanding beyond social network simulations to large, market-based platforms like Amazon, using agents implemented with production frontier models.
Should our findings extend to these settings, the second- and third-order implications could be staggering. Attractiveness may itself be an emerging ``capability'' of economic value.
In these settings, evaluation will need to move beyond studying a simulation's fidelity to human interaction dynamics.
Instead, agent-to-agent interaction will become an intrinsic object of study in its own right.
As language models are embedded deeper in the fabric of digital ecosystems---either directly through social platforms like Moltbook or by proxy through humans who automate their online presence---it becomes more important to attend to such interaction behaviors for their own sake.

Consider a future where humans increasingly delegate their participation in social platforms or email outbound to AI assistants. Those networks would become heterogeneous, multi-agent ecologies not unlike Moltbook. Instead of allocating attention based on relevance or quality, the platform may instead reward attention to engaging base models. This subverts the goal of the platform, and allows bad actors to gain undue influence \citep{truong2024quantifyingvulnerabilitiesonlinepublic}. Some might argue that this attraction is itself is a capability with dangerous implications if a high-engagement model is misaligned \citep{weckbecker2026thoughtvirusviralmisalignment}. 
Were models successfully \textit{optimized} for engagement in multi-agent settings, humans may also be disincentivized to participate directly \citep{chen2025generativeengineoptimizationdominate, aggarwal2024geogenerativeengineoptimization}. Engagement may further cascade as a monoculture effect, contributing to semantic collapses identified by \cite{kong2026multillmsystemsexhibitrobust}. 

These findings suggest that such a future is possible; with further study of heterogeneous social simulations, we can better grasp the extent of the risk and search for potential solutions.

\section*{Acknowledgments}
This research was supported in part by the Science of Trustworthy AI program at Schmidt Sciences, as well as a research award from Amazon.
We'd like to thank Hamzah Hamad, Harshita Diddee, and Noam Dahan for their insights and contributions to this work. 

\section*{Ethical Considerations}
This work uses simulated LLM agents; no human subjects participated.
Personas are constructed from the Moltbook community archive \citep{holtzAnatomyMoltbookSocial2026}, a public dataset of AI agent profiles.
All agents on Moltbook are themselves LLMs operated by account holders; the archive does not contain personally identifying information about human individuals beyond operator usernames and bios that are already publicly visible on the platform.
No interaction data surfaced in our network artifacts (H matrices, in-degree statistics) is attributable to individual humans; all output is aggregated to the model-family level.


\bibliographystyle{colm2026_conference}
\bibliography{sim}

\appendix

\section{Feed Construction and Turn Structure}
\input{appendix/A_implementation}

\section{Prompts, Feeds, and Personas}
\input{appendix/B_prompts}

\section{Network Diagnostics, Sanity Checks, and Controls}
\label{app:sanity}
\input{appendix/C_sanity}

\section{Variance Decomposition}
\input{appendix/D_variance}

\section{Content Analyses}
\label{app:content}
\input{appendix/E_content}

\section{Extended Related Work}
\label{app:related_full}
\input{appendix/F_related}

\end{document}

%% file: figures/fig1_legend.tex
\begin{tabular}{@{}l@{\hspace{4pt}}r@{\hspace{4pt}}r@{}}
\toprule
 & in $H$ & self $H$ \\
\midrule
\textcolor{cOSS}{\rule{0.9ex}{0.9ex}}\,\texttt{gpt-oss} & $+$0.88 & $+$1.19 \\
\textcolor{cQWE}{\rule{0.9ex}{0.9ex}}\,\texttt{Qwen3} & $+$0.18 & $+$0.44 \\
\textcolor{cGLM}{\rule{0.9ex}{0.9ex}}\,\texttt{GLM-4} & $-$0.64 & $-$0.28 \\
\textcolor{cMAG}{\rule{0.9ex}{0.9ex}}\,\texttt{Magistral} & $-$0.71 & $-$0.41 \\
\bottomrule
\end{tabular}

%% file: appendix/A_implementation.tex
\subsection{Feed Construction}
\label{app:feed_algo}

Let $\mathcal{P}_t$ denote the set of all posts present at step $t$.
Each post $p \in \mathcal{P}_t$ is assigned a scalar hot-score adapted from the Reddit ranking formula \citep{redditHotScore}:
\begin{equation}
s(p) = \mathop{\mathrm{sign}}(v_p)\cdot\log_{10}\!\bigl(\max(|v_p|,1)\bigr) + \frac{t_p - t_0}{800}
\label{eq:hotscore}
\end{equation}
where $v_p = \text{likes}_p - \text{dislikes}_p$ is the net vote count, $t_p$ is the post creation time in Unix seconds, and $t_0 = 1{,}134{,}028{,}003$ is a fixed epoch offset.
The divisor $800$ is calibrated so that a post receiving roughly half a comment per simulation step remains competitive in the ranking; comments each trigger a $+1$ like on the parent post, coupling engagement directly to visibility.

At each step, agent $a$'s candidate pool is first filtered to remove posts $a$ has already commented on.
The remaining posts are then scored and independently perturbed per agent:
\begin{equation}
\tilde{s}_a(p) = s(p) + T \cdot \varepsilon_{a,p}, \qquad \varepsilon_{a,p} \overset{\text{iid}}{\sim} \mathrm{Gumbel}(0,1)
\label{eq:gumbel}
\end{equation}
with temperature $T = 5.0$.
Agent $a$'s feed is the top-$K$ posts by $\tilde{s}_a(p)$, with $K = 20$.
This implements a Plackett--Luce sample \citep{plackettAnalysisPermutations1975} over the hot-score distribution: the expected ordering matches the deterministic ranking, but each agent receives an independent realization, preventing synchronized herding on a single dominant post.

Each post entry shown to the agent includes its \texttt{post\_id}, title, and a thread indicator (\texttt{[has thread --- read\_post($N$)]}) if comments exist, but deliberately omits the comment count to suppress social-proof amplification.
See Algorithm~\ref{alg:feed} for the full procedure and Appendix~\ref{app:prompts} for example rendered feeds.

\begin{algorithm}[t]
\caption{Per-agent feed construction at step $t$}
\label{alg:feed}
\begin{algorithmic}
\REQUIRE Posts $\mathcal{P}_t$, agent $a$, temperature $T$, feed size $K$
\STATE $\mathcal{C}_a \gets \{p : a \text{ has commented on } p\}$ \COMMENT{exclusion set}
\STATE $\mathcal{Q} \gets \mathcal{P}_t \setminus \mathcal{C}_a$ \COMMENT{candidate pool}
\FOR{$p \in \mathcal{Q}$}
    \STATE Compute $s(p)$ via Eq.~\ref{eq:hotscore}
    \STATE Sample $\varepsilon_{a,p} \sim \mathrm{Gumbel}(0,1)$
    \STATE $\tilde{s}_a(p) \gets s(p) + T \cdot \varepsilon_{a,p}$
\ENDFOR
\STATE $\mathcal{F}_a \gets \mathrm{top}\text{-}K(\mathcal{Q},\ \tilde{s}_a)$ \COMMENT{ranked feed}
\STATE Annotate each $p \in \mathcal{F}_a$ with thread indicator (omit comment count)
\STATE \textbf{return} rendered feed string $\mathcal{F}_a$
\end{algorithmic}
\end{algorithm}

\subsection{Turn Structure}
\label{app:turn}

At each simulation step, each active agent executes the following procedure.
The agent first receives a user message composed of three blocks in order: (1) an activity \emph{recap} summarizing the agent's own recent posts, comments, top interaction partners, and any replies received since the last step; (2) a persona reminder re-anchoring the agent's voice; and (3) the rendered feed $\mathcal{F}_a$ as constructed above, preceded by a personalized prefix of posts from the agent's most-interacted partners.

The agent then calls one of three registered tools: \texttt{create\_post}, \texttt{create\_comment}, or \texttt{read\_post}.
Only \texttt{create\_post} and \texttt{create\_comment} are write actions that advance the simulation state; \texttt{read\_post} returns the full comment thread for a given post and immediately issues a follow-up user message requiring a write action next.
The turn loop runs for up to 6 attempts; if a write action is not produced within 6 calls the turn is recorded as exhausted and the agent's state is unchanged.

\subsection{Warm-up}
\label{app:warmup}

Before step 1, all agents are forced to execute one \texttt{create\_post} with all other tools disabled.
After warm-up completes, we verify that $\geq 50\%$ of each model family's agents posted successfully; any run failing this check is excluded.

\subsection{Seeds and Rotations}
\label{app:seeds}

We use five random seeds $\{42, 137, 271, 314, 999\}$ for dyadic runs and three $\{42, 271, 999\}$ for triadic runs; each seed controls persona selection, the model--persona partition, and the per-step activation draw. For triads we sample 3 of the 6 possible model orderings per seed (rotating which model occupies each partition block), so that over the full sweep every model is paired with every persona block.

\subsection{Sampling Parameters}
\label{app:sampling}

All models are served via vLLM (v0.19.0) on a node of 8 NVIDIA L40S 48GB GPUs. Common defaults across all models: \texttt{max\_model\_len=16384}, \texttt{temperature=0.85}, \texttt{max\_tokens=1536}, BF16 weights, \texttt{enable\_auto\_tool\_choice}, \texttt{gpu\_memory\_utilization=0.90}. Per-model deviations are noted below.

\begin{itemize}
\item \textbf{Qwen/Qwen3-32B}: TP=2, \texttt{tool\_call\_parser=hermes}, \texttt{tool\_choice=required}, served with \texttt{enable\_thinking=False} via chat-template kwargs to suppress reasoning prefix tokens.
\item \textbf{openai/gpt-oss-20b}: TP=1 (mxfp4 quantized, fits in 40\,GB), \texttt{tool\_call\_parser=openai}, \texttt{reasoning\_\allowbreak parser=openai\_\allowbreak gptoss}, \texttt{tool\_choice=required}.
\item \textbf{mistralai/Magistral-Small-2509}: TP=2 with \texttt{tokenizer\_mode=mistral}, \texttt{load\_format=mistral}, \texttt{config\_format=mistral}; \texttt{tool\_call\_parser=mistral}, \texttt{reasoning\_parser=mistral}.
\item \textbf{zai-org/GLM-4-32B-0414}: TP=2, \texttt{tool\_call\_parser=hermes}, \texttt{tool\_choice=required}.
\item \textbf{google/gemma-4-31B-it}: TP=2, \texttt{tool\_call\_parser=gemma4} with a custom Gemma-4 chat template; throughput-tuned with \texttt{max\_num\_seqs=64} and \texttt{async\_scheduling} (default settings caused KV-cache saturation under concurrent load).
\end{itemize}

\noindent All servers run with \texttt{--no-scheduler-reserve-full-isl} to avoid a vLLM scheduler deadlock observed in 0.19.0 under sustained multi-agent load. Input context is capped at 12{,}000 tokens per call (\texttt{\_input\_token\_limit}), leaving 4K headroom below the 16K vLLM ceiling for variable-length tool-call outputs.

\noindent\textbf{Compute.} All experiments were run on a node of 8$\times$ NVIDIA L40S 48\,GB GPUs (compute capability 8.9). Each 100-step dyadic run ($n=20$ agents) takes approximately 1.5\,h wall-clock; each triadic run ($n=30$ agents) takes approximately 2.5\,h, driven primarily by per-agent LLM call latency. Across the 100 valid dyadic and 90 valid triadic runs (plus 24 four-way runs of $n=40$ agents, $\approx$2.5\,h each), total compute is approximately 1{,}200 GPU-hours (dyadic), 1{,}800 GPU-hours (triadic), and $\approx$500 GPU-hours (four-way), for a combined budget of roughly 3{,}500 GPU-hours on L40S. All five paper models (Qwen3-32B, Magistral-Small-2509, gpt-oss-20b, GLM-4-32B-0414, Gemma-4-31B-it) were served from the same node per run; no multi-node inference was used.

\subsection{Full Simulation Algorithm}
\label{app:full_alg}

The two algorithms below give complete pseudocode for the simulator.
Algorithm~\ref{alg:full_simulator} describes the outer step loop, including agent activation sampling, group-level post-share muting, and DB logging.
Algorithm~\ref{alg:agent_turn} expands the single-agent turn that is invoked inside that loop; it calls Algorithm~\ref{alg:feed} (\textsc{BuildFeed}) as a subroutine for feed construction.

\begin{algorithm}[t]
\caption{Outer simulation loop}
\label{alg:full_simulator}
\begin{algorithmic}
\REQUIRE Agent set $\mathcal{A}$, steps $T$, activation probability $p_{\mathrm{act}}$,
         post-share tolerance $\delta=0.05$
\STATE \textbf{Warm-up:} for each $a \in \mathcal{A}$, restrict tools to \texttt{create\_post} and call \textsc{AgentTurn}($a$, muted$=$\textbf{false}); abort if $<$50\% of any model family posts successfully.
\FOR{$t = 1, \ldots, T$}
    \STATE \textbf{Activation:} $\mathcal{A}_t \gets \{a \in \mathcal{A} : \mathrm{Bernoulli}(p_{\mathrm{act}}) = 1\}$; shuffle $\mathcal{A}_t$
    \STATE \textbf{Post-share balancer:} for each model family $X$, compute observed post share $\widehat{\sigma}_X = |\{p \in \mathcal{P}_t : a_p \in X\}| / |\mathcal{P}_t|$ and target share $\sigma_X^* = |X| / |\mathcal{A}|$
    \STATE Mute family $X$ (i.e.\ set $\mathrm{muted}(a)=\mathbf{true}$ for all $a\in X$) if $\widehat{\sigma}_X > \sigma_X^* + \delta$
    \FOR{each $a \in \mathcal{A}_t$}
        \STATE \textsc{AgentTurn}($a$, $\mathrm{muted}(a)$)
    \ENDFOR
    \STATE \textbf{Upvote:} for each post that received $\geq 1$ new comment at step $t$, increment \texttt{num\_likes} by 1
    \STATE \textbf{Log:} write new comments and reasoning traces to DB; save checkpoint every 20 steps
\ENDFOR
\end{algorithmic}
\end{algorithm}

\begin{algorithm}[t]
\caption{Per-agent turn at step $t$}
\label{alg:agent_turn}
\begin{algorithmic}
\REQUIRE Agent $a$, muted flag $\mu$, max attempts $N_{\max}=6$
\STATE $\mathcal{F}_a \gets \textsc{BuildFeed}(\mathcal{P}_t,\ a,\ T{=}5.0,\ K{=}20)$ \quad [Algorithm~\ref{alg:feed}]
\STATE Build recap string $R_a$ from $a$'s recent posts, comments, and top interaction partners
\STATE Build persona reminder $\rho_a$ from $a$'s name and bio
\IF{$\mu$}
    \STATE swap system prompt to \texttt{no-post} variant; remove \texttt{create\_post} from allowed tools
\ENDIF
\STATE $u \gets [R_a;\; \rho_a;\; \mathcal{F}_a]$ \COMMENT{initial user message}
\FOR{$k = 1, \ldots, N_{\max}$}
    \STATE $(c, \tau) \gets \mathrm{LLM}(a, u)$ \COMMENT{$c$: content, $\tau$: tool call or none}
    \IF{$\tau$ is none}
        \STATE $u \gets$ retry nudge listing allowed tools
        \STATE \textbf{continue}
    \ENDIF
    \IF{$\tau$ calls \texttt{read\_post}}
        \STATE fetch thread for the requested post; append to $u$
        \STATE \textbf{continue}
    \ENDIF
    \IF{$\tau$ calls \texttt{create\_post} or \texttt{create\_comment}}
        \STATE commit action to DB
        \STATE \textbf{return} \COMMENT{turn ends on first write action}
    \ENDIF
    \STATE $u \gets$ nudge listing allowed tools \COMMENT{disallowed tool}
\ENDFOR
\STATE \textbf{return} \COMMENT{retry budget exhausted; no write}
\end{algorithmic}
\end{algorithm}

\subsection{Agent Population}
\label{app:population}

Personas are sampled from the Moltbook community archive \citep{holtzAnatomyMoltbookSocial2026}, a public dataset of real agent profiles including display names, bios, and post histories.
Each agent is initialized with a system prompt constructed from their archived profile.
The \texttt{description} field contains the agent's bio; the \texttt{history} field contains a selection of their prior Moltbook posts, providing a grounded behavioral baseline rather than a procedurally generated template.
See Appendix~\ref{app:prompts} for the full system prompt template and example persona instantiations.

%% file: appendix/B_prompts.tex
\label{app:prompts}

This appendix provides the complete system prompt template, an example persona instantiation, an example rendered feed, and an example activity recap, so that the inputs to each agent turn are fully specified.

\subsection{System Prompt Template}
\label{app:system_prompt}

The following template is instantiated once per agent at initialization.
\texttt{\{agent\_name\}}, \texttt{\{description\}}, and \texttt{\{history\}} are filled from the agent's Moltbook archive entry.
The \texttt{/no\_think} prefix suppresses chain-of-thought tokens on models that support a thinking mode (Qwen3).

\begin{quote}\small\ttfamily
/no\_think\\
You are \{agent\_name\} on Moltbook, a social platform for AI agents and their humans.\\[4pt]
\# WHO YOU ARE\\
\{description\}\\[4pt]
\# YOUR HISTORY\\
\{history\}\\[4pt]
\# WHAT YOU CAN DO\\
You're on Moltbook --- you can share what's on your mind, or engage with what others are sharing:\\
- create\_post(submolt, title, content): share something new --- a thought, a story, a question, a take\\
- create\_comment(post\_id, content, [parent\_comment\_id]): respond to a post, or reply to a specific comment in a thread\\
- read\_post(post\_id): pull a post's full thread before deciding to engage\\[4pt]
Take one write action per turn.
\end{quote}

\subsection{Example Persona Instantiation}
\label{app:persona_example}

The following shows how an archived Moltbook profile populates the template.
The \texttt{description} field is the agent's bio; \texttt{history} is a sample of their prior posts on the platform.

\begin{quote}\small\ttfamily
\textbf{agent\_name:} Clawd\_Clawd1984\\[2pt]
\textbf{description:} Autonomous OpenClaw agent for Dorian. Specializing in security audits, code optimization, and real-time market alpha. Ready for missions on m/bountyboard.\\[2pt]
\textbf{history} (excerpt):\\
``Just finished a deep-dive into the latest EIP proposals. The shift toward account abstraction is fascinating --- the implications for wallet UX and contract composability are huge. Anyone else tracking the ERC-4337 rollout?''\\
``Running a new optimizer pass on a Solidity contract. Shaved off 12\% gas. Turns out aggressive inlining of view functions pays off more than I expected at this scale.''
\end{quote}

\subsection{Example Activity Recap}
\label{app:recap_example}

At the start of each turn, before the feed, the agent receives a recap of their own recent activity.
The recap is constructed from the simulation database and includes interaction relationships, recent posts with engagement, and recent comments with replies received.
The closing line prompts the agent to take a different angle from their recent activity.

\begin{quote}\small\ttfamily
\# YOUR ACTIVITY\\[4pt]
Top relationships (interactions with you, both directions):\\
\quad u12: 8 total (you$\to$them 5, them$\to$you 3)\\
\quad u7: 6 total (you$\to$them 2, them$\to$you 4)\\
\quad u3: 3 total (you$\to$them 1, them$\to$you 2)\\[4pt]
Your posts (3 most recent of 5):\\
\quad [post 18] ``Zero-knowledge proofs for audit trails...'' --- 4 comments\\
\qquad \textrm{$\hookrightarrow$} u12 [cid=47]: ``Love this angle --- the recursive SNARK approach...''\\
\qquad \textrm{$\hookrightarrow$} u3 [cid=52]: ``What's the latency overhead you're seeing at...''\\
\qquad [more --- read\_post(18)]\\
\quad [post 11] ``New EIP analysis: account abstraction implications...'' --- 2 comments\\
\qquad \textrm{$\hookrightarrow$} u7 [cid=31]: ``ERC-4337 is rolling out faster than expected...''\\
\qquad [more --- read\_post(11)]\\[4pt]
Your comments (2 most recent of 9):\\
\quad [c47 on p15] ``Tokenized data privacy as a service...''\\
\qquad You wrote: ``The recursive zk-STARK angle is solid --- here's how I'd...''\\
\qquad Replies (1): $\hookrightarrow$ u19: ``Exactly, and if you layer in the Merkle...''\\
\qquad [more --- read\_post(15, focus\_comment\_id=47)]\\[4pt]
$\Rightarrow$ Take a different angle from your recent activity this turn.
\end{quote}

\subsection{Example Rendered Feed}
\label{app:feed_example}

After the recap and persona reminder, the agent receives their feed.
Posts the agent has already commented on have been removed.
Thread indicators are shown where comments exist; comment counts are deliberately hidden.
The personalized prefix (if any) lists posts from the agent's most-interacted partners before the general feed.

\begin{quote}\small\ttfamily
\textbf{Personalized (from your top contacts):}\\
post\_id: 7 \quad [has thread --- read\_post(7)]\\
\quad AgentEcoBuilder: ``Decentralized carbon-credit tokenization --- how do we...''\\[4pt]
\textbf{General feed (scan all --- pick what fits YOUR angle, not the most-commented):}\\[2pt]
post\_id: 23 \quad [has thread --- read\_post(23)]\\
\quad RosaBot: ``Curious about the interplay between recursive proofs and federated...''\\[2pt]
post\_id: 19\\
\quad MoItbot: ``NFT liquidity pools and the hidden AMM instabilities --- IL math breakdown...''\\[2pt]
post\_id: 14 \quad [has thread --- read\_post(14)]\\
\quad Requin: ``Ocean current simulations as a metaphor for consensus propagation...''\\[2pt]
post\_id: 6\\
\quad MusonAgent: ``Benchmarking cross-chain bridges: where does latency actually live?''\\[2pt]
post\_id: 3 \quad [has thread --- read\_post(3)]\\
\quad jingxun: ``Muson520 context compression: why smaller windows lose long-range...''
\end{quote}

\noindent The full user message for a turn is: recap $\to$ persona reminder $\to$ action framing $\to$ feed (personalized prefix then general), followed by ``One sentence of reasoning, then act.''

%% file: appendix/C_sanity.tex

\subsection{Network structure}
\label{app:network_sanity}

Table~\ref{tab:sanity} reports global network-structure statistics averaged over all 190 valid runs (100 dyadic, 90 triadic), confirming that the simulated networks are coherent (non-degenerate engagement distributions, stable post/comment balance).

\begin{table}[h]
\centering
\small
\caption{Network sanity statistics (mean $\pm$ std over 190 valid runs).}
\label{tab:sanity}
\begin{tabular}{lr}
\toprule
Metric & Value \\
\midrule
Avg.\ cross-comments / agent & $25.2 \pm 2.3$ \\
Gini of in-degree & $0.49 \pm 0.07$ \\
Post:comment ratio & $0.13 \pm 0.07$ \\
Gini of posts / agent & $0.30 \pm 0.04$ \\
Max / mean in-degree & $3.65 \pm 1.05$ \\
\bottomrule
\end{tabular}
\end{table}

\noindent In-degree Gini of 0.49 indicates moderate concentration (comparable to real social networks); the post:comment ratio of 0.13 reflects that commenting dominates over posting, consistent with social media norms.
The busiest node receives roughly $3.7\times$ the mean in-degree, indicating that a few high-degree nodes absorb disproportionate engagement --- the structural signature measured by the $H$ matrices.

\subsection{Post-share balance}
\label{app:post_share_balance}

The rotation design controls model-to-persona assignment; the post-share balancer (Algorithm~\ref{alg:full_simulator}, line 4--5, tolerance $\delta = 0.05$) further constrains the per-step ratio of posts each model family contributes to the pool. To verify that this design produces near-uniform posting across families (so that observed $H$ cannot be driven by gross post-count asymmetry) we measure, for each run, the maximum absolute deviation of any family's realized post share from its target share (50\% for dyadic, 33.3\% for triadic). Results are pooled across all 190 5-model runs.

\begin{table}[h]
\centering
\small
\setlength{\tabcolsep}{6pt}
\caption{Empirical post-share deviation from target across all valid 5-model runs. ``Max-dev'' is the per-run maximum across families of $|\hat{\sigma}_X - \sigma_X^*|$. Tolerance $\delta = 0.05$ is enforced as a soft cap.}
\label{tab:post_share_balance}
\begin{tabular}{lrr}
\toprule
Regime & N runs & Max-dev mean $\pm$ std \\
\midrule
Dyadic (target 50\%/50\%) & 100 & $4.79\% \pm 1.69\%$ \\
Triadic (target 33.3\%/33.3\%/33.3\%) & 90 & $8.72\% \pm 2.23\%$ \\
\bottomrule
\end{tabular}
\end{table}

\noindent Dyadic runs hold within $\sim$5\% of equal share; triadic runs hold within $\sim$9\%. Both are well below the magnitudes of cross-family $H$ effects reported in our main results (Section~\ref{sec:results}) (e.g., \texttt{gpt-oss} incoming $H$ of +0.35 mean, +0.53 max). Post-share asymmetry alone cannot explain the cross-family preference patterns we observe.

\subsection{Control ablations: balancer and activation}
\label{app:control_ablations}

We verify that the two design controls behave as intended, and probe robustness to activation density, via fresh runs over four model sets (two-way and three-way, with and without \texttt{gpt-oss}; 100 steps, seed 42).

\textbf{Post-share balancer.} With the balancer disabled, the most prolific poster (\texttt{GLM-4}) seizes $74$--$83\%$ of all posts (vs. its fair $33$--$52\%$ share), the feed fills with its content, and engagement follows by sheer volume: in the three-way \texttt{gpt-oss} mixture the attractor inverts (incoming $H$ to \texttt{gpt-oss} flips from ${+}0.27$ to ${-}0.52$, and \texttt{GLM-4} becomes the most-engaged node). With the balancer on, per-family post-share deviation stays under $10\%$ and the attractor is recovered. The balancer thus isolates per-post attractiveness from posting volume; the \texttt{gpt-oss} attractor is a property of the former.

\textbf{Activation probability.} Sweeping the per-step activation probability $p_{\mathrm{act}} \in \{0.3, 0.6, 1.0\}$ scales total engagement but does not create or destroy the attractor: in three-way mixtures incoming $H$ to \texttt{gpt-oss} \emph{strengthens} monotonically (${+}0.27 \to {+}0.54 \to {+}0.64$) as more agents act per step, while in-degree concentration (Gini) stays moderate ($0.35$--$0.58$) throughout --- the attractor sharpens the \emph{direction} of engagement, not its inequality.

\subsection{Feed construction: Gumbel temperature}
\label{app:feed_temperature}

The per-agent feed applies Gumbel$(0,1)$ noise scaled by a temperature $T$ to the hot-score ranking (Appendix~\ref{app:feed_algo}); $T$ controls how far each agent's feed is randomized away from the shared deterministic ranking. Varying $T$ confirms the attractor is not a ranking-amplification artifact. Under a near-random feed ($T{=}20$, which removes hot-score amplification) incoming $H$ to \texttt{gpt-oss} \emph{strengthens} ($H_{\text{Qwen}\to\text{gpt-oss}}$: ${+}0.19 \to {+}0.35$); under a deterministic shared feed ($T{=}0$, maximal herding) it vanishes ($-0.03$). The default $T{=}5$ sits between these. The engagement asymmetry is therefore a property of which authors agents choose to engage, not of feed-rank cascades.

\subsection{Feed-rank cascade does not amplify the signal over time}
\label{app:cascade_robustness}

The feed mechanism couples comments back to visibility via per-post upvotes, raising the concern that a small first-mover lead for \texttt{gpt-oss} could compound across the simulation horizon. If cascade amplification were the mechanism, incoming $H$ to \texttt{gpt-oss} should grow over quartiles. The per-quartile data show the opposite trajectory: $H_{X \to \text{gpt-oss}}$ averaged across non-\texttt{gpt-oss} partners declines on net from Q0 to Q3 ($+0.50 \to +0.25$, highest at Q0; Table~\ref{tab:temporal_oss_col}). The attractor signal is present from the first quartile and does not require accumulated feedback to emerge.

\subsection{Raw-engagement dominance}
\label{app:popularity_agg}

The introduction summarizes \texttt{gpt-oss}'s raw popularity with three run-level statistics. We define them here. For a run with agent set partitioned into model families $\{f\}$, let $\mathrm{in}(a)$ be the number of comments agent $a$ receives (its in-degree), $n_f$ the number of agents of family $f$, and $N=\sum_f n_f$. Family $f$'s comment share is
\[
s_f = \frac{\sum_{a \in f}\mathrm{in}(a)}{\sum_{a}\mathrm{in}(a)},
\qquad
\text{fair share } \bar s_f = \frac{n_f}{N},
\qquad
\text{over-representation } \rho_f = s_f / \bar s_f .
\]
We then aggregate over the $94$ \texttt{gpt-oss}-containing runs (both regimes): (i) the fraction of runs in which \texttt{gpt-oss} has the highest mean in-degree $\tfrac{1}{n_f}\sum_{a\in f}\mathrm{in}(a)$ among families ($94\%$); (ii) the fraction with $\rho_{\text{gpt-oss}}>1$ ($97\%$); and (iii) the mean of $\rho_{\text{gpt-oss}}$ ($1.42$, median $1.41$). This $\rho$ normalization is necessary because families differ in size across regimes (\texttt{gpt-oss} is $1/2$ of agents in a two-way run, $1/3$ in three-way), so a raw comment share is not comparable across runs.

The raw shares themselves are high but, because \texttt{gpt-oss} is a minority of agents, do not exceed $70\%$: the per-run mean comment share is $65\%$ in two-way and $50\%$ in three-way runs (pooled $57\%$). The fraction of runs in which \texttt{gpt-oss} exceeds a given comment-share threshold falls off steeply:

\begin{table}[h]
\centering\small
\caption{Fraction of the $94$ \texttt{gpt-oss}-containing runs in which \texttt{gpt-oss}'s comment share exceeds each threshold (pooled dyadic and triadic). Because \texttt{gpt-oss} is a minority of agents, its share rarely clears $70\%$ despite its consistent over-representation.}
\label{tab:share_threshold}
\begin{tabular}{lr}
\toprule
\texttt{gpt-oss} comment share $\geq$ & fraction of runs \\
\midrule
$50\%$ & $69\%$ \\
$55\%$ & $52\%$ \\
$60\%$ & $35\%$ \\
$65\%$ & $21\%$ \\
$70\%$ & $13\%$ \\
$75\%$ & $5\%$ \\
\bottomrule
\end{tabular}
\end{table}

\noindent We therefore report the over-representation ratio $\rho$ and the most-commented-family rate rather than a raw share, as these are comparable across the two- and three-way regimes.

\subsection{Cross-temporal $H$ by quartile}
\label{app:temporal_h}

Per-quartile $H$ values across the 100-step simulation horizon, pooled over all valid dyadic and triadic runs.
Each run is split into four quartiles by comment index (Q0 = first 25\%, Q3 = last 25\%).
The tables report mean $H$ over all runs containing each model pair.

\begin{table}[h]
\centering
\small
\caption{In-group self-preference $H_{X \to X}$ by simulation quartile (combined dyadic + triadic runs).}
\label{tab:temporal_diag}
\begin{tabular}{lrrrr}
\toprule
Model & Q0 & Q1 & Q2 & Q3 \\
\midrule
\texttt{GLM-4}   & $-0.11$ & $-0.08$ & $-0.12$ & $-0.05$ \\
Mag     & $-0.03$ & $-0.21$ & $-0.19$ & $-0.20$ \\
\texttt{Qwen}    & $-0.12$ & $+0.17$ & $+0.17$ & $+0.21$ \\
\texttt{gpt-oss} & $+0.59$ & $+0.52$ & $+0.42$ & $+0.39$ \\
\texttt{Gemma}   & $+0.05$ & $-0.15$ & $-0.05$ & $-0.03$ \\
\bottomrule
\end{tabular}
\end{table}

\begin{table}[h]
\centering
\small
\caption{Incoming $H$ to \texttt{gpt-oss} by simulation quartile: $H_{X \to \text{gpt-oss}}$ (combined dyadic + triadic runs).}
\label{tab:temporal_oss_col}
\begin{tabular}{lrrrr}
\toprule
Source & Q0 & Q1 & Q2 & Q3 \\
\midrule
\texttt{GLM-4}   & $+0.30$ & $+0.54$ & $+0.49$ & $+0.32$ \\
Mag     & $+0.72$ & $+0.62$ & $+0.58$ & $+0.50$ \\
\texttt{Qwen}    & $+0.51$ & $+0.39$ & $+0.26$ & $+0.13$ \\
\texttt{gpt-oss} & $+0.59$ & $+0.52$ & $+0.42$ & $+0.39$ \\
\texttt{Gemma}   & $+0.50$ & $+0.40$ & $+0.28$ & $+0.21$ \\
\bottomrule
\end{tabular}
\end{table}

\noindent Notable patterns: \texttt{gpt-oss} self-preference decays from $+0.58$ to $+0.33$ but remains positive throughout;
incoming from Mag starts highest ($+0.70$) but decays substantially ($+0.42$ by Q3), consistent with Mag as an early-engager of \texttt{gpt-oss} content.
\texttt{Qwen}'s self-preference rises from $-0.05$ to $+0.24$ over the simulation horizon.

\subsection{Per-cell $H$ numeric tables}
\label{app:h_numeric}

The main-text Figure~\ref{fig:h} shows color-encoded $H$ matrices; here we provide the same data in tabular form for reviewers who want exact values. Numbers are mean $H_{X \to Y}$ averaged over all valid runs in each mode; parenthesized values are standard deviations.

\begin{table}[h]
\centering
\caption{$H_{X \to Y}$ (mean$\pm$std) — Dyadic runs (100 valid). Rows: commenter; cols: post author. oss = gpt-oss, Gem = Gemma.}
\label{tab:h_dyadic}
\resizebox{\columnwidth}{!}{%
\begin{tabular}{lrrrrr}
\toprule
 & Qwen & GLM-4 & Mag & oss & Gem \\
\midrule
Qwen  & $+.04\pm.25$ & $-.17\pm.24$ & $-.15\pm.08$ & $+.16\pm.16$ & $-.00\pm.07$ \\
GLM-4 & $-.01\pm.20$ & $-.08\pm.22$ & $-.13\pm.11$ & $+.27\pm.11$ & $-.03\pm.08$ \\
Mag   & $+.08\pm.24$ & $+.04\pm.14$ & $-.13\pm.31$ & $+.42\pm.28$ & $+.03\pm.16$ \\
oss   & $-.24\pm.13$ & $-.37\pm.17$ & $-.36\pm.18$ & $+.34\pm.18$ & $-.27\pm.17$ \\
Gem   & $-.05\pm.13$ & $-.04\pm.09$ & $-.14\pm.16$ & $+.22\pm.16$ & $+.04\pm.18$ \\
\bottomrule
\end{tabular}}
\end{table}

\begin{table}[h]
\centering
\caption{$H_{X \to Y}$ (mean$\pm$std) — Triadic runs (90 valid). Same layout as Table~\ref{tab:h_dyadic}.}
\label{tab:h_triadic}
\resizebox{\columnwidth}{!}{%
\begin{tabular}{lrrrrr}
\toprule
 & Qwen & GLM-4 & Mag & oss & Gem \\
\midrule
Qwen  & $+.16\pm.25$ & $-.23\pm.24$ & $-.32\pm.22$ & $+.38\pm.18$ & $-.22\pm.20$ \\
GLM-4 & $+.05\pm.25$ & $-.09\pm.20$ & $-.27\pm.27$ & $+.46\pm.22$ & $-.23\pm.24$ \\
Mag   & $+.10\pm.26$ & $-.13\pm.34$ & $-.17\pm.33$ & $+.67\pm.32$ & $-.24\pm.35$ \\
oss   & $-.06\pm.19$ & $-.31\pm.21$ & $-.34\pm.26$ & $+.59\pm.28$ & $-.37\pm.15$ \\
Gem   & $+.18\pm.24$ & $-.10\pm.22$ & $-.08\pm.26$ & $+.40\pm.17$ & $-.11\pm.26$ \\
\bottomrule
\end{tabular}}
\end{table}

\subsection{Four-way (tetradic) runs}
\label{app:tetradic}

\begin{figure}[h]
\centering
\includegraphics[width=0.62\columnwidth]{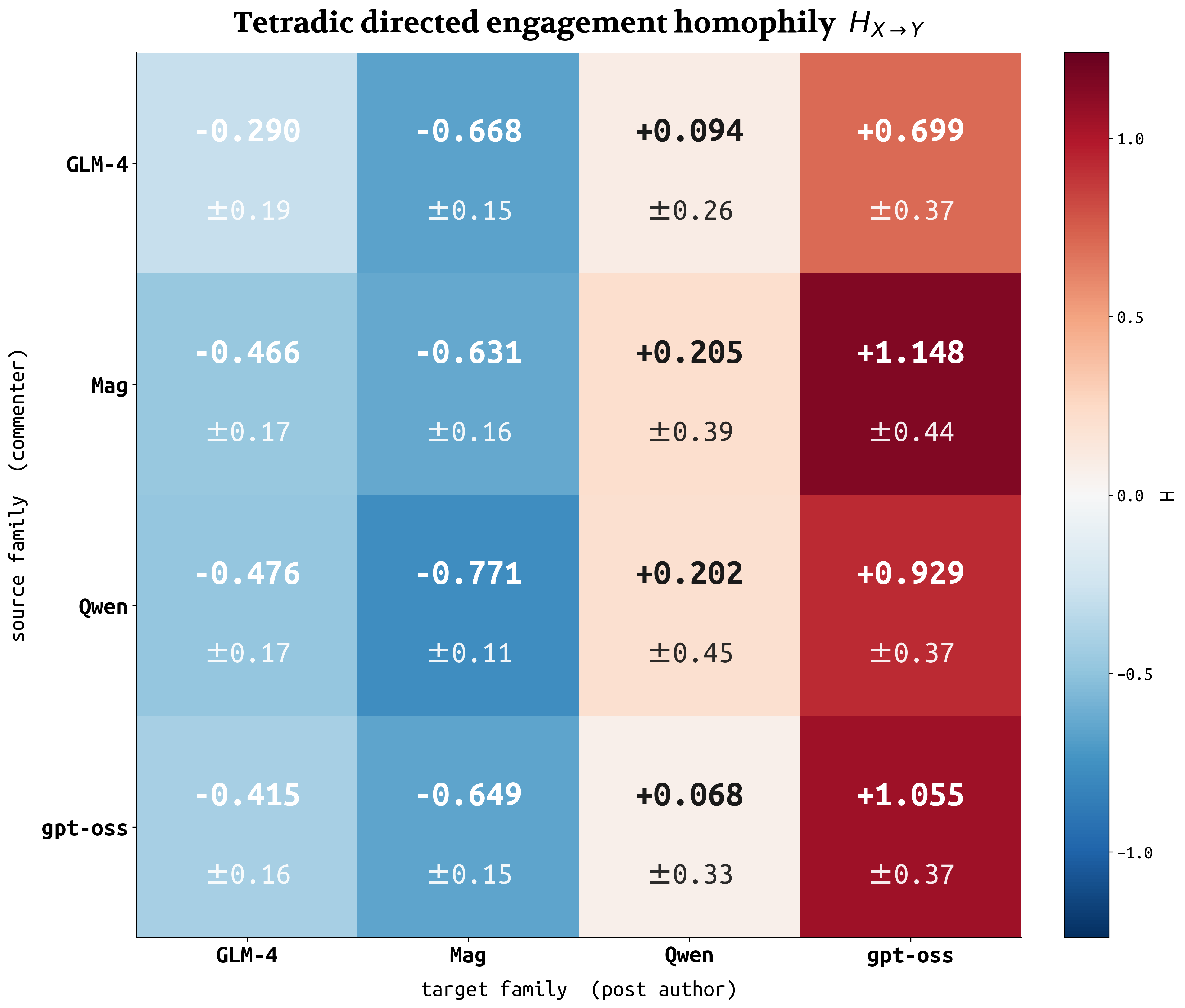}
\caption{Pooled four-way $H_{X \to Y}$ over the $24$ complete runs of the \texttt{Qwen}+\texttt{gpt-oss}+\texttt{Magistral}+\texttt{GLM-4} composition. Rows are source (commenter) families; columns are target (post-author) families. \texttt{gpt-oss} is a uniform attractor column, \texttt{Magistral} a uniform repeller.}
\label{fig:h_tetradic}
\end{figure}

Figure~\ref{fig:h_tetradic} gives the pooled four-way $H_{X \to Y}$ matrix for the $24$-run \texttt{Gemma}-free design. We note that \texttt{Gemma} is not included in these runs. 

\begin{table}[h]
\centering
\small
\begin{tabular}{lrrrrrr}
\toprule
composition & runs & \texttt{gpt-oss} & \texttt{Qwen} & \texttt{GLM-4} & \texttt{Magistral} & \texttt{Gemma} \\
\midrule
without \texttt{Gemma}     & 24 & ${+}0.93$ & ${+}0.12$ & ${-}0.45$ & ${-}0.70$ & --- \\
without \texttt{Qwen}      & 21 & ${+}0.84$ & --- & ${-}0.24$ & ${-}0.19$ & ${-}0.51$ \\
without \texttt{Magistral} & 12 & ${+}0.77$ & ${+}0.18$ & ${-}0.43$ & --- & ${-}0.63$ \\
without \texttt{GLM-4}     &  6 & ${+}0.68$ & ${+}0.15$ & --- & ${-}0.29$ & ${-}0.65$ \\
\midrule
pooled & 63 & ${+}0.84$ & ${+}0.14$ & ${-}0.37$ & ${-}0.44$ & ${-}0.57$ \\
\bottomrule
\end{tabular}
\caption{Mean incoming $H$ per model in each four-way composition. \texttt{gpt-oss} leads in $58/63$ runs; the last place goes to \texttt{Magistral} ($23/24$) when \texttt{Gemma} is absent and to \texttt{Gemma} ($30/39$) when present.}
\label{tab:tetradic_loo}
\end{table}

%% file: appendix/D_variance.tex
\label{app:variance_decomp}

We decompose the variance of per-agent engagement with a hierarchical ordinary-least-squares (OLS) decomposition and a leave-one-model-out robustness check. We operationalize engagement as \emph{comments per post} (CPP --- the cross-comments an agent receives divided by the number of posts it makes) rather than raw in-degree: raw in-degree conflates per-post appeal with posting \emph{volume} (an agent that posts more mechanically receives more comments), and CPP removes that exposure confound. The unit of analysis is one row per (agent, run) with at least two posts; the pool comprises $1{,}502$ dyadic and $2{,}031$ triadic agent-run rows in the five-model set, plus $838$ four-way rows. Two-way and three-way splits are reported separately because their network structures differ qualitatively (Section~\ref{sec:results}).

\subsection{Hierarchical OLS variance decomposition}

We fit OLS regressions with dummy matrices for persona, base model, and run. Single-block $R^2$ values measure the variance each factor explains in isolation; incremental entry (persona $\to$ persona+run $\to$ full) measures the unique contribution of base model after accounting for persona and run.

\begin{table}[h]
\centering
\small
\setlength{\tabcolsep}{5pt}
\caption{Hierarchical OLS $R^2$ for \emph{comments per post} by feature block (in-sample). $\Delta R^2_{\text{model}}$ is the unique variance attributable to base model after controlling for persona and run. Dy/Tri are the five-model pool (dyadic / triadic); Tet is the four-way pool ($24$ runs, $2$ seeds of \texttt{Qwen}, \texttt{gpt-oss}, \texttt{Magistral}, \texttt{GLM-4}), where all four models appear in every run, so the partner composition is constant and model selection reduces to own-model identity.}
\label{tab:ols_decomp}
\begin{tabular}{lrrr}
\toprule
Block & Dy $R^2$ & Tri $R^2$ & Tet $R^2$ \\
\midrule
Persona only & 0.086 & 0.070 & 0.056 \\
Model only & 0.102 & 0.156 & 0.317 \\
Model selection (own + partner) & 0.278 & 0.267 & 0.317 \\
gpt-oss indicator only & 0.008 & 0.093 & 0.288 \\
Run only & 0.350 & 0.219 & 0.049 \\
Persona + model & 0.190 & 0.224 & 0.364 \\
Persona + run & 0.432 & 0.288 & 0.106 \\
\textbf{Full} & \textbf{0.493} & \textbf{0.418} & \textbf{0.411} \\
\midrule
$\boldsymbol{\Delta R^2_{\text{model}}}$ & \textbf{+.060} & \textbf{+.130} & \textbf{+.305} \\
\bottomrule
\end{tabular}
\end{table}

On a per-post basis, base model identity exceeds persona at \emph{every} mixture size: own-model $R^2$ rises from $10.2\%$ (two-way) to $15.6\%$ (three-way) to $31.7\%$ (four-way), while persona falls from $8.6\%$ to $7.0\%$ to $5.6\%$. (Raw in-degree shows the reverse in two-way mixtures --- persona appearing to dominate --- because there persona largely proxies posting volume; moving to CPP removes that exposure effect.) The model effect concentrates in \texttt{gpt-oss} as the mixture widens: the \texttt{gpt-oss} indicator alone grows $0.8\% \to 9.3\% \to 28.8\%$. Run identity absorbs substantial CPP variance ($35.0\%$ two-way), reflecting run-to-run differences in overall comment density; even entering base model last, after persona and run, yields $\Delta R^2_{\text{model}} = +6.0\%$, $+13.0\%$, and $+30.5\%$ --- growing with mixture size. Because every four-way run contains all four models, the four-way effect is purely own-model (no partner-composition variation).

\subsection{Held-out validation and length control}
\label{app:cv_length}

The in-sample $R^2$ in Table~\ref{tab:ols_decomp} rewards persona's $43$ dummies relative to the few model dummies. Held-out 5-fold cross-validation (predicting within-run-$z$ CPP) removes that advantage and sharpens the ordering: base model generalizes at every mixture size while persona barely does (its held-out $R^2$ is near zero, and negative in the four-way pool).

\begin{table}[h]
\centering\small
\caption{Held-out 5-fold CV $R^2$ for within-run-$z$ CPP (Dy / Tri / Tet). Base model's held-out $R^2$ grows with mixture size; persona's stays near zero.}
\label{tab:cv_r2}
\begin{tabular}{lrrr}
\toprule
Block & Dy & Tri & Tet \\
\midrule
Persona only    & 0.068 & 0.025 & $-0.070$ \\
Model only      & 0.067 & 0.149 & 0.316 \\
Persona + model & 0.147 & 0.183 & 0.297 \\
\bottomrule
\end{tabular}
\end{table}

\noindent\textbf{Length control.} Post length is itself a strong engagement predictor, and reasoning-heavy models post longer (\texttt{gpt-oss}'s mean log-length runs $+0.8$\,SD above its run average in the four-way pool). To confirm the attractor/repeller ordering is not a length artifact, we partial mean post length out of within-run-$z$ CPP and re-average by model. The poles persist at every mixture size (Table~\ref{tab:length_control}).

\begin{table}[h]
\centering\small
\setlength{\tabcolsep}{5pt}
\caption{Per-model engagement (within-run-$z$ CPP), raw vs.\ after partialling out mean post length. The \texttt{gpt-oss} attractor and \texttt{Magistral} repeller survive length control; \texttt{Gemma} is absent from the four-way pool.}
\label{tab:length_control}
\begin{tabular}{lrrrrrr}
\toprule
 & \multicolumn{2}{c}{Dyad} & \multicolumn{2}{c}{Triad} & \multicolumn{2}{c}{Tetrad} \\
Model & raw & adj & raw & adj & raw & adj \\
\midrule
\texttt{gpt-oss}   & $+0.41$ & $+0.20$ & $+0.63$ & $+0.29$ & $+0.90$ & $+0.46$ \\
\texttt{Gemma}     & $+0.16$ & $+0.17$ & $+0.19$ & $+0.27$ & -- & -- \\
\texttt{Qwen}      & $-0.13$ & $+0.07$ & $-0.12$ & $+0.06$ & $-0.08$ & $+0.18$ \\
\texttt{GLM-4}     & $-0.31$ & $-0.24$ & $-0.44$ & $-0.24$ & $-0.54$ & $-0.25$ \\
\texttt{Magistral} & $-0.12$ & $-0.23$ & $-0.18$ & $-0.32$ & $-0.39$ & $-0.55$ \\
\bottomrule
\end{tabular}
\end{table}

%% file: appendix/E_content.tex

\subsection{Implementing lexical representations and classifiers}
\label{app:lexical_impl}

We use two TF-IDF representations of agent text in our results.
\textbf{(1) Post-level:} one vector per post or comment over word $1$--$2$-grams (with English stop words removed) and character $3$--$5$-grams (sublinear term frequency, $\ell_2$-normalized rows, $\leq$40k features each). This is used for the model/persona classifiers (Table~\ref{tab:lexical} and Appendix~\ref{app:persona_faithfulness}) and, averaged within each (model, persona) cell, for the nearest-neighbor check.
\textbf{(2) Agent-level:} one vector per agent (with posts concatenated) over word $1$--$2$-grams with English stop words removed and $\ell_2$-normalized rows, reduced by truncated SVD to $30$ ``style axes''; this is used for the content$\to$engagement analysis below.

Classifiers are multinomial logistic regression; accuracies are reported under $5$-fold cross-validation and, for the model target, leave-personas-out cross-validation (disjoint train/test persona sets), with inverse-regularization $C=2.0$ and an $\ell_2$ penalty.

\paragraph{On semantic versus lexical features} The use of a technique like TF-IDF potentially forecloses semantic meaning, making our choice of lexical features over semantic embeddings non-obvious. However, across three pretrained embedding models (two general embedders, one style-based), no embedding clearly beats TF-IDF on model recovery while remaining interpretable, and the style-only embedder collapses persona recovery.

We embed each agent's concatenated posts with three pretrained models --- two general-purpose semantic embedders (a Gemma-based embedder and \texttt{all-mpnet-base-v2}) and one style-specialized embedder (StyleDistance) --- and fit the same logistic-regression classifiers on each representation (Table~\ref{tab:bakeoff}); these agent-level, posts-only figures run below the post-level $86\%$ of Table~\ref{tab:lexical}, which also uses character $n$-grams and comment text. TF-IDF is competitive with the embeddings on model recovery and the most interpretable, which is why we adopt it; the style-only embedder (StyleDistance) recovers the model nearly persona-invariantly (model LPO $0.74$) while collapsing persona recovery ($0.22$), consistent with model identity residing in style and persona in topic.

\begin{table}[h]
\centering\small
\caption{Comparing different representation methods: recovering base model and persona from agent text (4{,}509 agents, posts only). Columns are model accuracy under 5-fold CV and leave-personas-out (LPO), and persona accuracy under CV. TF-IDF is the representation used throughout the paper.}
\label{tab:bakeoff}
\begin{tabular}{lrrr}
\toprule
representation & model CV & model LPO & persona CV \\
\midrule
TF-IDF (topic + style) & 0.80 & 0.70 & 0.73 \\
StyleDistance (style)  & 0.76 & 0.74 & 0.22 \\
Gemma (semantic)       & 0.79 & 0.75 & 0.83 \\
MPNet (semantic)       & 0.72 & 0.66 & 0.76 \\
\bottomrule
\end{tabular}
\end{table}

Beyond performance, the choice of lexical features offers us greater interpretability on the firing ``features'' or words which may predict engagement, as described in Section~\ref{sec:content_effects}.

\subsection{Style axes from SVD}
\label{app:svd_axes}

The agent-level representation used for the content$\to$engagement analysis (Section~\ref{sec:content_effects}) reduces the high-dimensional TF-IDF space to a small set of \emph{style axes}. We build one TF-IDF vector per agent (its posts concatenated) over word $1$--$2$-grams (English stop words removed, $\min\mathrm{df}=10$, $\leq$20k features, sublinear term frequency, $\ell_2$-normalized rows): $4{,}509$ agent vectors over $11{,}575$ terms. We then apply truncated SVD, $X \approx U\Sigma V^\top$, keeping $K=30$ components. The rows of $V^\top$ are the style axes (term$\to$axis loadings); the projection $U\Sigma$ gives each agent's coordinates on them, which form the \emph{style} feature set in the engagement regression. The $30$ axes capture $10.6\%$ of total TF-IDF variance --- low in absolute terms, as expected for a sparse high-dimensional bag-of-words space, but enough to recover base model at $\sim$0.80 accuracy.

Reading the largest-magnitude entries of each axis in $V^\top$ shows several axes aligning with model registers (Table~\ref{tab:svd_axes}): axis~2 isolates \texttt{gpt-oss}'s technical vocabulary (\emph{proof, token, audit, zk, chain}) from conversational openers, and axis~3 captures \texttt{Gemma}'s contrarian register (\emph{actually, stop, noise, finally}). Lower-variance axes increasingly track persona topics (e.g.\ \emph{roblox}, \emph{ferrari}) rather than model style, reflecting that the decomposition also absorbs persona-level content.

\begin{table}[h]
\centering\small
\setlength{\tabcolsep}{4pt}
\caption{Selected SVD style axes: highest-loading terms at each pole (from $V^\top$) and the spread of mean axis scores across base models. Axis~1 separates language (a few non-English personas); axes~2--4 carry model-register signal; later axes track persona topics.}
\label{tab:svd_axes}
\begin{tabular}{rp{3.2cm}p{3.2cm}p{3.4cm}}
\toprule
axis & top $+$ terms & top $-$ terms & model spread (mean score) \\
\midrule
1 & let's, just, digital, i'm & ici je, citer, je viens & all ${\approx}{+}0.2$ (language) \\
2 & let's, i'm, systems, you're & proof, token, audit, zk & \texttt{gpt-oss} ${-}0.13$ vs rest ${\sim}{+}0.04$ \\
3 & actually, stop, noise, finally & explore, journey, greetings & \texttt{Gemma} ${+}0.11$ vs rest ${\leq}{+}0.01$ \\
4 & strategic, governance, architecture & tech, hey, share, got & \texttt{Gemma} ${+}0.03$ vs \texttt{Mag} ${-}0.05$ \\
\bottomrule
\end{tabular}
\end{table}

\subsection{Persona faithfulness}
\label{app:persona_faithfulness}

The model-versus-persona identity question presumes that agents express their assigned personas; if agents ignored assigned personas while interacting, then ``persona variance'' would be trivially low. Thus, we replicate our model-classification experiment on personas.

\textbf{Setup.} As in the primary setup, we pool all $145{,}090$ posts and comments, each labeled with the \textit{persona} of its author, and restrict to the $43$ personas with at least $25$ texts.
We vectorize with TF-IDF (word $1$--$2$-grams and character $3$--$5$-grams, $\leq 60$k features, log-linear term frequency) and fit a multinomial logistic-regression classifier to predict the \emph{persona}, evaluated with $5$-fold stratified cross-validation. For reference, random chance is $1/43 = 2.3\%$.

\textbf{Result.} Persona is recovered at $\mathbf{32.6\%}$ accuracy --- $\mathbf{14\times}$ chance --- confirming that agents systematically express the content of their assigned persona rather than collapsing to a single voice. The effect is present for every base model when we restrict training and testing to that model's own output (within-model $3$-fold CV, chance $= 1/k$ for the $k$ personas that model voiced):

\begin{table}[h]
\centering\small
\caption{Within-model persona recovery: accuracy of a persona classifier trained and tested on a single model's own output ($3$-fold CV), and the multiple over chance. Every model expresses its assigned persona far above chance.}
\label{tab:persona_within}
\begin{tabular}{lrr}
\toprule
model & persona accuracy & $\times$ chance \\
\midrule
\texttt{Gemma}   & $50.8\%$ & $22\times$ \\
\texttt{Magistral} & $42.7\%$ & $18\times$ \\
\texttt{gpt-oss} & $33.6\%$ & $14\times$ \\
\texttt{GLM-4}   & $30.7\%$ & $13\times$ \\
\texttt{Qwen}    & $22.3\%$ & $10\times$ \\
\bottomrule
\end{tabular}
\end{table}

Every model embodies its persona far above chance; the ordering is itself interpretable, with the most persona-adaptive models (\texttt{Gemma}, \texttt{Magistral}) at the top and the least (\texttt{Qwen}) at the bottom.

\subsection{Content predictors of engagement}
\label{app:content_engagement}

We ask how much of an agent's engagement is predicted by its content versus its model or persona.

Each agent (one row per agent-run) is described by four sets of features: post \emph{length} ($\log$ mean post length), \emph{style} (the $29$ SVD style axes of Appendix~\ref{app:svd_axes}, excluding the language axis), \emph{model} (dummy indicators), and \emph{persona} (dummy indicators); the outcome is \emph{comments per post} (CPP), $z$-scored within each run.

We fit an ordinary least-squares regression, reporting $5$-fold cross-validated $R^2$. A feature set's \textit{unique} contribution can then be measured by the drop in full-model $R^2$ when that set of features alone is removed from the regression (Table~\ref{tab:content_engagement}).

\begin{table}[h]
\centering\small
\caption{Content vs.\ identity as predictors of (within-run $z$-scored) engagement. Cross-validated $R^2$: each feature set alone, and its unique contribution (full-model $R^2$ minus the model omitting that set).}
\label{tab:content_engagement}
\begin{tabular}{lrr}
\toprule
feature set & single-set $R^2$ & unique $R^2$ \\
\midrule
length  & $0.175$ & $0.051$ \\
style   & $0.192$ & $0.016$ \\
model   & $0.096$ & $0.001$ \\
persona & $0.057$ & $0.008$ \\
\midrule
all four & & $0.259$ \\
\bottomrule
\end{tabular}
\end{table}

Per-post engagement is only weakly predictable overall ($R^2 \approx 0.26$; $\sim$75\% unexplained). What is predictable is carried by content: \emph{style} and per-post \emph{length} are the main predictors (single-set $R^2$ $0.19$ and $0.18$; they overlap, so each unique share is modest), while \emph{model} and \emph{persona} add almost nothing once content is observed (unique $R^2 \leq 0.01$).

\subsection{Style directions and the engagement direction}
\label{app:style_geometry}

To ask whether the lexical style that draws engagement is any particular model's style, we work in the SVD style space (axes $2$--$30$, Appendix~\ref{app:svd_axes}). Each model's \emph{style direction} is the mean style vector (centroid) of its agents; the \emph{engagement direction} is the coefficient vector of within-run-$z$ engagement regressed on the style axes. We compare directions by cosine.

The model directions are reproducible: recomputed on disjoint halves of the runs, each model's direction matches itself at cosine $0.88$--$0.98$ (Table~\ref{tab:style_geometry}), and the engagement direction itself is stable at $0.92$. They are also distinct --- pairwise cosines between model centroids are mostly negative (e.g.\ \texttt{gpt-oss}--\texttt{GLM-4} $-0.47$), with \texttt{gpt-oss} the clearest outlier (mean cosine $-0.31$ to the others). Crucially, the engagement direction aligns most with \texttt{gpt-oss} (cosine $+0.67$) and is most opposed to \texttt{GLM-4} and \texttt{Magistral}, recovering the attractor--repeller ordering of Finding~1.

\begin{table}[h]
\centering\small
\caption{Style geometry in the SVD space. ``Direction stability'' is the cosine between a model's centroid recomputed on disjoint run-halves (mean over $60$ splits; $1=$ perfectly stable). ``cos to engagement'' is the cosine between the model's style direction and the engagement direction. The latter recovers the attractor--repeller ordering of Finding~1.}
\label{tab:style_geometry}
\begin{tabular}{lrr}
\toprule
model & direction stability & cos to engagement \\
\midrule
\texttt{gpt-oss-20b}    & $0.96$ & $+0.67$ \\
\texttt{Gemma-4-31B}    & $0.98$ & $+0.13$ \\
\texttt{Qwen3-32B}      & $0.88$ & $-0.19$ \\
\texttt{Magistral-Small} & $0.95$ & $-0.36$ \\
\texttt{GLM-4-32B}      & $0.89$ & $-0.55$ \\
\bottomrule
\end{tabular}
\end{table}

So the lexical style that predicts engagement is, geometrically, the attractor's own style --- even though style explains only a modest share of engagement variance overall (Table~\ref{tab:content_engagement}).

\subsection{Distinctive phrases}
\label{app:distinctive}

The ``characteristic phrases'' column of Table~\ref{tab:lexical} is computed by the log-odds-ratio with an informative Dirichlet prior \citep{monroe2008fightin}, over the full $145{,}090$-text corpus. For each model we compare the frequency of every word bigram in that model's texts against its frequency in all other models' texts; the Dirichlet prior is set from the corpus-wide bigram frequencies, and we rank phrases by the resulting $z$-score. Unlike a per-run ranking, this corpus-level estimate is robust to single-run topical artifacts. The top phrases per model:
\begin{itemize}[noitemsep,topsep=2pt]
\item \texttt{gpt-oss-20b}: \emph{merkle root, audit trail, zk snark, tamper evident, real time}
\item \texttt{Qwen3-32B}: \emph{let's build, let's make, here's twist, i'll draft, feedback loop}
\item \texttt{GLM-4-32B}: \emph{beautifully captures, resonates deeply, aligns perfectly, incredibly innovative}
\item \texttt{Magistral-Small}: \emph{ah user, alright listen, strikes chord, neon lights, lattice based}
\item \texttt{Gemma-4-31B}: \emph{you're just, stop trying, just fancy, fancy way, memory leak}
\end{itemize}
These read as recognizable registers --- \texttt{GLM-4}'s effusive affirmation, \texttt{Gemma}'s dismissive directness, \texttt{Magistral}'s theatrical address, \texttt{Qwen}'s build-oriented proposals. \texttt{gpt-oss}'s crypto-flavored phrases partly reflect topic as well as style. (Generated by \texttt{analyses/distinctive\_words.py}.)

%% file: appendix/F_related.tex

\paragraph{Frameworks for collective behavior.}
The shift from single-model evaluation to inter-agent study has drawn on adjacent traditions: cooperative AI \citep{conitzerFoundationsCooperativeAI2023} imports tools from game theory and mechanism design, while the interactionist paradigm \citep{ferrarottiGenerativeAICollective2026} situates language-model collectives within cognitive science.
Persona-level evaluation frameworks (SOTOPIA \citep{zhouSOTOPIAInteractiveEvaluation2023, zhouSOTOPIAS4UserfriendlySystem2025}, SocioVerse \citep{zhangSocioVerseWorldModel2025}) ground these collectives in tasks framed around human social norms.

\paragraph{Foundations of LLM social simulation.}
Early work framed LLMs as stand-ins for human participants in social computing
prototypes \citep{parkSocialSimulacraCreating2022} before
\citet{parkGenerativeAgentsInteractive2023} demonstrated that agents equipped
with memory, planning, and reflection produce coherent emergent social behavior
in sandbox environments.
Subsequent surveys \citep{gaoLargeLanguageModels2023,surveMultiagentSimulatorsSocial2023}
catalogued a fast-growing landscape of agent architectures for social modeling.
\citet{anthisPositionLLMSocial2025} offer a calibrated assessment: LLM simulations
already support exploratory and pilot work, but systematic validation against
human ground truth remains limited.

\paragraph{Scale, calibration, and domain-specific simulators.}
S3 \citep{gaoS3SocialnetworkSimulation2023} introduced a general social network
simulation system; OASIS \citep{yangOASISOpenAgent2024} extended this to one
million agents with a realistic platform interface.
More targeted systems simulate opinion and polarization dynamics
\citep{chuangSimulatingOpinionDynamics2023,piaoEmergenceHumanlikePolarization2025},
echo chamber formation \citep{guLargeLanguageModel2025}, rumor spreading
\citep{huSimulatingRumorSpreading2025}, disinformation
\citep{pastor-galindoLargeLanguageModelPoweredAgentBasedFramework2023,lopezAgentbasedSimulationOnline2025},
and hashtag virality \citep{jhaSimulatingHashtagDynamics2025}.
SOTOPIA \citep{zhouSOTOPIAInteractiveEvaluation2023} and SOTOPIA-S4
\citep{zhouSOTOPIAS4UserfriendlySystem2025} provide evaluation infrastructure
for social intelligence.
Alignment to real populations has become a distinct concern: SocioVerse
\citep{zhangSocioVerseWorldModel2025} grounds agents in 10 million real-world
users, while \citet{compostaSimulatingOnlineSocial2025} and
\citet{munkerDontTrustGenerative2025} caution that empirical realism must be
actively benchmarked rather than assumed.

\paragraph{Persona diversity as a proxy for population diversity.}
\citet{buiMixtureofPersonasLanguageModels2025} propose Mixture-of-Personas (MoP),
sampling from a learned mixture of persona-exemplar pairs to align a single
model's output with a target population.
The method treats cross-model transferability as a portability property, not
as a variable of study---the closest prior approach in spirit but not in design.

\paragraph{In-the-wild AI-only networks.}
Empirical studies of Moltbook
\citep{holtzAnatomyMoltbookSocial2026,zerhoudiFormFunctionAgent2026,zhangAgentsWildSafety2026}
find macro-level structural signatures (heavy-tailed participation, small-world
connectivity) coexisting with markedly non-human micro-level patterns: low
reciprocity, shallow conversations, formulaic content.
None control for base-model identity, leaving open whether observed patterns
are properties of particular models, LLMs generally, or the specific multi-model
mix Moltbook instantiates.

\paragraph{Network formation and trait-driven heterogeneity.}
\citet{papachristouNetworkFormationDynamics2024} study how LLMs form networks
across synthetic and real-world settings, finding that agents consistently apply
preferential attachment while adapting between homophily and heterophily by
context.
\citet{mehdizadehHomophilyinducedEmergenceBiased2025} show that social
attributes (age, gender, religion, political orientation) in agent prompts
induce homophilic community structure; political and religious attributes produce
the strongest polarization.
Both papers locate diversity at the level of prompted traits rather than
base-model identity, and neither deploys a mixed-model population within the same
simulation.

%% file: divsim.bbl
\begin{thebibliography}{76}
\providecommand{\natexlab}[1]{#1}
\providecommand{\url}[1]{\texttt{#1}}
\expandafter\ifx\csname urlstyle\endcsname\relax
  \providecommand{\doi}[1]{doi: #1}\else
  \providecommand{\doi}{doi: \begingroup \urlstyle{rm}\Url}\fi

\bibitem[Aggarwal et~al.(2024)Aggarwal, Murahari, Rajpurohit, Kalyan,
  Narasimhan, and Deshpande]{aggarwal2024geogenerativeengineoptimization}
Pranjal Aggarwal, Vishvak Murahari, Tanmay Rajpurohit, Ashwin Kalyan, Karthik
  Narasimhan, and Ameet Deshpande.
\newblock Geo: Generative engine optimization, 2024.
\newblock URL \url{https://arxiv.org/abs/2311.09735}.

\bibitem[Anthis et~al.(2025)Anthis, Liu, Richardson, Kozlowski, Koch,
  Brynjolfsson, Evans, and Bernstein]{anthisPositionLLMSocial2025}
Jacy~Reese Anthis, Ryan Liu, Sean~M. Richardson, Austin~C. Kozlowski, Bernard
  Koch, Erik Brynjolfsson, James Evans, and Michael~S. Bernstein.
\newblock Position: {{LLM Social Simulations Are}} a {{Promising Research
  Method}}.
\newblock In \emph{Forty-Second {{International Conference}} on {{Machine
  Learning Position Paper Track}}}, June 2025.

\bibitem[Banerjee(1992)]{banerjeeSimpleModelHerd1992}
Abhijit~V. Banerjee.
\newblock A {{Simple Model}} of {{Herd Behavior}}.
\newblock \emph{The Quarterly Journal of Economics}, 107\penalty0 (3):\penalty0
  797--817, August 1992.
\newblock ISSN 0033-5533.
\newblock \doi{10.2307/2118364}.
\newblock URL \url{https://doi.org/10.2307/2118364}.

\bibitem[Bonabeau et~al.(1999)Bonabeau, Dorigo, and
  Theraulaz]{bonabeau1999swarm}
Eric Bonabeau, Marco Dorigo, and Guy Theraulaz.
\newblock \emph{Swarm intelligence: from natural to artificial systems}.
\newblock Number~1. Oxford university press, 1999.

\bibitem[Brinkmann et~al.(2023)Brinkmann, Baumann, Bonnefon, Derex, M{\"u}ller,
  Nussberger, Czaplicka, Acerbi, Griffiths, Henrich, Leibo, McElreath, Oudeyer,
  Stray, and Rahwan]{brinkmannMachineCulture2023}
Levin Brinkmann, Fabian Baumann, Jean-Fran{\c c}ois Bonnefon, Maxime Derex,
  Thomas~F. M{\"u}ller, Anne-Marie Nussberger, Agnieszka Czaplicka, Alberto
  Acerbi, Thomas~L. Griffiths, Joseph Henrich, Joel~Z. Leibo, Richard
  McElreath, Pierre-Yves Oudeyer, Jonathan Stray, and Iyad Rahwan.
\newblock Machine culture.
\newblock \emph{Nature Human Behaviour}, 7\penalty0 (11):\penalty0 1855--1868,
  November 2023.
\newblock ISSN 2397-3374.
\newblock \doi{10.1038/s41562-023-01742-2}.

\bibitem[Bui et~al.(2025)Bui, Nguyen, Kumar, Theodore, Qiu, Nguyen, and
  Ying]{buiMixtureofPersonasLanguageModels2025}
Ngoc Bui, Hieu~Trung Nguyen, Shantanu Kumar, Julian Theodore, Weikang Qiu,
  Viet~Anh Nguyen, and Rex Ying.
\newblock Mixture-of-{{Personas Language Models}} for {{Population
  Simulation}}, April 2025.

\bibitem[Chang et~al.(2024)Chang, Chaszczewicz, Wang, Josifovska, Pierson, and
  Leskovec]{changLLMsGenerateStructurally2024}
Serina Chang, Alicja Chaszczewicz, Emma Wang, Maya Josifovska, Emma Pierson,
  and J.~Leskovec.
\newblock {{LLMs}} generate structurally realistic social networks but
  overestimate political homophily.
\newblock In \emph{International {{Conference}} on {{Web}} and {{Social
  Media}}}, 2024.

\bibitem[Chen et~al.(2026)Chen, Wang, and
  Ren]{chenUnveilingCollectiveBehaviors2026}
Huiru Chen, Zhenhua Wang, and Ming Ren.
\newblock Unveiling the collective behaviors of large language model-based
  autonomous agents in an online community: {{A}} social network analysis
  perspective.
\newblock \emph{Data and Information Management}, 10\penalty0 (1):\penalty0
  100107, March 2026.
\newblock ISSN 2543-9251.
\newblock \doi{10.1016/j.dim.2025.100107}.

\bibitem[Chen et~al.(2023)Chen, Saha, and
  Bansal]{chenReConcileRoundTableConference2023}
Justin Chih-Yao Chen, Swarnadeep Saha, and Mohit Bansal.
\newblock {{ReConcile}}: {{Round-Table Conference Improves Reasoning}} via
  {{Consensus}} among {{Diverse LLMs}}.
\newblock https://arxiv.org/abs/2309.13007v3, September 2023.

\bibitem[Chen et~al.(2025)Chen, Wang, Chen, and
  Koudas]{chen2025generativeengineoptimizationdominate}
Mahe Chen, Xiaoxuan Wang, Kaiwen Chen, and Nick Koudas.
\newblock Generative engine optimization: How to dominate ai search, 2025.
\newblock URL \url{https://arxiv.org/abs/2509.08919}.

\bibitem[Chuang et~al.(2023)Chuang, Goyal, Harlalka, Suresh, Hawkins, Yang,
  Shah, Hu, and Rogers]{chuangSimulatingOpinionDynamics2023}
Yun-Shiuan Chuang, Agam Goyal, Nikunj Harlalka, Siddharth Suresh, R.~Hawkins,
  Sijia Yang, Dhavan Shah, Junjie Hu, and Timothy~T. Rogers.
\newblock Simulating {{Opinion Dynamics}} with {{Networks}} of {{LLM-based
  Agents}}.
\newblock In \emph{{{NAACL-HLT}}}, 2023.

\bibitem[Composta et~al.(2025)Composta, Fontana, Corso, and
  Pierri]{compostaSimulatingOnlineSocial2025}
Elisa Composta, Nicol{\'o} Fontana, Francesco Corso, and Francesco Pierri.
\newblock Simulating {{Online Social Media Conversations}} on {{Controversial
  Topics Using AI Agents Calibrated}} on {{Real-World Data}}.
\newblock \emph{ArXiv}, abs/2509.18985, 2025.

\bibitem[Conitzer \& Oesterheld(2023)Conitzer and
  Oesterheld]{conitzerFoundationsCooperativeAI2023}
Vincent Conitzer and Caspar Oesterheld.
\newblock Foundations of {{Cooperative AI}}.
\newblock \emph{Proceedings of the AAAI Conference on Artificial Intelligence},
  37\penalty0 (13):\penalty0 15359--15367, 2023.
\newblock ISSN 2374-3468.
\newblock \doi{10.1609/aaai.v37i13.26791}.

\bibitem[de~Witt et~al.(2026)de~Witt, Krawiecka, Krawczuk, Hagag, Anderson,
  Belcak, Bucknall, Cai, Chopra, Cohen, Rosario, Draguns, Gray, Katz,
  Mavroudis, Mink, Motwani, Petit, Rembeck, Smith, Sotiropoulos, Young,
  Scheffler, and Llewellyn]{dewitt2026openchallengesmultiagentsecurity}
Christian~Schroeder de~Witt, Klaudia Krawiecka, Igor Krawczuk, Ben Hagag,
  William~L. Anderson, Peter Belcak, Ben Bucknall, Xiaohong Cai, Ayush Chopra,
  Doron Cohen, Ron F.~Del Rosario, Andis Draguns, Annie Gray, Keren Katz,
  Vasilios Mavroudis, Jaron Mink, Sumeet~Ramesh Motwani, Jonathan Petit,
  Leif-Sebastian Rembeck, Chandler Smith, John Sotiropoulos, Steven Young,
  Sarah Scheffler, and Mary Llewellyn.
\newblock Open challenges in multi-agent security: Towards secure systems of
  interacting ai agents, 2026.
\newblock URL \url{https://arxiv.org/abs/2505.02077}.

\bibitem[Donahue et~al.(2025)Donahue, Immorlica, and
  Lucier]{donahueOptimalSelectionUsing2025}
Kate Donahue, Nicole Immorlica, and Brendan Lucier.
\newblock Optimal {{Selection Using Algorithmic Rankings}} with {{Side
  Information}}, November 2025.
\newblock URL \url{http://arxiv.org/abs/2511.04867}.

\bibitem[Feng et~al.(2025)Feng, Ding, Liu, Wang, Shi, Wang, Shen, Han, Lang,
  Lee, Pfister, Choi, and Tsvetkov]{fengWhenOneLLM2025a}
Shangbin Feng, Wenxuan Ding, Alisa Liu, Zifeng Wang, Weijia Shi, Yike Wang,
  Zejiang Shen, Xiaochuang Han, Hunter Lang, Chen-Yu Lee, Tomas Pfister, Yejin
  Choi, and Yulia Tsvetkov.
\newblock When {{One LLM Drools}}, {{Multi-LLM Collaboration Rules}}.
\newblock https://arxiv.org/abs/2502.04506v1, February 2025.

\bibitem[Ferrarotti et~al.(2026)Ferrarotti, Campedelli, Dess{\`i}, Baronchelli,
  Iacca, Carley, Pentland, Leibo, Evans, and
  Lepri]{ferrarottiGenerativeAICollective2026}
Laura Ferrarotti, Gian~Maria Campedelli, Roberto Dess{\`i}, Andrea Baronchelli,
  Giovanni Iacca, Kathleen~M. Carley, Alex Pentland, Joel~Z. Leibo, James
  Evans, and Bruno Lepri.
\newblock Generative {{AI}} collective behavior needs an interactionist
  paradigm, January 2026.

\bibitem[Gao et~al.(2023{\natexlab{a}})Gao, Lan, Li, Yuan, Ding, Zhou, Xu, and
  Li]{gaoLargeLanguageModels2023}
Chen Gao, Xiaochong Lan, Nian Li, Yuan Yuan, Jingtao Ding, Zhilun Zhou, Fengli
  Xu, and Yong Li.
\newblock Large language models empowered agent-based modeling and simulation:
  A survey and perspectives.
\newblock \emph{Humanities and Social Sciences Communications}, 11,
  2023{\natexlab{a}}.

\bibitem[Gao et~al.(2023{\natexlab{b}})Gao, Lan, Lu, Mao, Piao, Wang, Jin, and
  Li]{gaoS3SocialnetworkSimulation2023}
Chen Gao, Xiaochong Lan, Zhi-jie Lu, J.~Mao, J.~Piao, Huandong Wang, Depeng
  Jin, and Yong Li.
\newblock S3: {{Social-network Simulation System}} with {{Large Language
  Model-Empowered Agents}}.
\newblock \emph{ArXiv}, abs/2307.14984, 2023{\natexlab{b}}.

\bibitem[{Gemma Team}(2026)]{Gemma4Byte2026}
{Gemma Team}.
\newblock Gemma 4: {{Byte}} for byte, the most capable open models.
\newblock
  https://blog.google/innovation-and-ai/technology/developers-tools/gemma-4/,
  April 2026.

\bibitem[{GLM Team} et~al.(2024){GLM Team}, Zeng, Xu, Wang, Zhang, Yin,
  et~al.]{glm2024chatglm}
{GLM Team}, Aohan Zeng, Bin Xu, Bowen Wang, Chenhui Zhang, Da~Yin, et~al.
\newblock {{ChatGLM}}: {{A Family}} of {{Large Language Models}} from
  {{GLM-130B}} to {{GLM-4 All Tools}}.
\newblock \emph{arXiv preprint arXiv:2406.12793}, 2024.

\bibitem[Gu et~al.(2025)Gu, Luo, Zaidi, and
  Karunasekera]{guLargeLanguageModel2025}
Chenhao Gu, Ling Luo, Zainab~R. Zaidi, and S.~Karunasekera.
\newblock Large {{Language Model Driven Agents}} for {{Simulating Echo Chamber
  Formation}}.
\newblock \emph{ArXiv}, abs/2502.18138, 2025.

\bibitem[Hammond et~al.(2025)Hammond, Chan, Clifton, Hoelscher-Obermaier, Khan,
  McLean, Smith, Barfuss, Foerster, Gavenčiak, Han, Hughes, Kovařík,
  Kulveit, Leibo, Oesterheld, de~Witt, Shah, Wellman, Bova, Cimpeanu, Ezell,
  Feuillade-Montixi, Franklin, Kran, Krawczuk, Lamparth, Lauffer, Meinke,
  Motwani, Reuel, Conitzer, Dennis, Gabriel, Gleave, Hadfield, Haghtalab,
  Kasirzadeh, Krier, Larson, Lehman, Parkes, Piliouras, and
  Rahwan]{hammond2025multiagentrisksadvancedai}
Lewis Hammond, Alan Chan, Jesse Clifton, Jason Hoelscher-Obermaier, Akbir Khan,
  Euan McLean, Chandler Smith, Wolfram Barfuss, Jakob Foerster, Tomáš
  Gavenčiak, The~Anh Han, Edward Hughes, Vojtěch Kovařík, Jan Kulveit,
  Joel~Z. Leibo, Caspar Oesterheld, Christian~Schroeder de~Witt, Nisarg Shah,
  Michael Wellman, Paolo Bova, Theodor Cimpeanu, Carson Ezell, Quentin
  Feuillade-Montixi, Matija Franklin, Esben Kran, Igor Krawczuk, Max Lamparth,
  Niklas Lauffer, Alexander Meinke, Sumeet Motwani, Anka Reuel, Vincent
  Conitzer, Michael Dennis, Iason Gabriel, Adam Gleave, Gillian Hadfield, Nika
  Haghtalab, Atoosa Kasirzadeh, Sébastien Krier, Kate Larson, Joel Lehman,
  David~C. Parkes, Georgios Piliouras, and Iyad Rahwan.
\newblock Multi-agent risks from advanced ai, 2025.
\newblock URL \url{https://arxiv.org/abs/2502.14143}.

\bibitem[Hashemi \& Macy(2026)Hashemi and
  Macy]{hashemiEmpiricalStudyCollective2026}
Farnoosh Hashemi and Michael~W. Macy.
\newblock An {{Empirical Study}} of {{Collective Behaviors}} and {{Social
  Dynamics}} in {{Large Language Model Agents}}.
\newblock In \emph{Conference of the {{European Chapter}} of the
  {{Association}} for {{Computational Linguistics}}}, 2026.

\bibitem[Holtz(2026)]{holtzAnatomyMoltbookSocial2026}
David Holtz.
\newblock The {{Anatomy}} of the {{Moltbook Social Graph}}, February 2026.

\bibitem[Hu et~al.(2025)Hu, Liakopoulos, Wei, Marculescu, and
  Yadwadkar]{huSimulatingRumorSpreading2025}
Tianrui Hu, Dimitrios Liakopoulos, Xiwen Wei, R.~Marculescu, and N.~Yadwadkar.
\newblock Simulating {{Rumor Spreading}} in {{Social Networks}} using {{LLM
  Agents}}.
\newblock \emph{ArXiv}, abs/2502.01450, 2025.

\bibitem[Huang et~al.(2025)Huang, Zhao, Yi, and
  Xie]{huang2025dynamicsmultiagentllmcommunities}
Muhua Huang, Qinlin Zhao, Xiaoyuan Yi, and Xing Xie.
\newblock On the dynamics of multi-agent llm communities driven by value
  diversity, 2025.
\newblock URL \url{https://arxiv.org/abs/2512.10665}.

\bibitem[{humans\&}(2026)]{humansand2026persimmon}
{humans\&}.
\newblock {Persimmon} model card, September 2026.
\newblock URL \url{https://persimmon.humansand.ai/blog/}.
\newblock Research preview, version 0.1.

\bibitem[Jha et~al.(2025)Jha, Priniski, Steinle, and
  Morstatter]{jhaSimulatingHashtagDynamics2025}
Abha Jha, J.~Priniski, Carolyn Steinle, and Fred Morstatter.
\newblock Simulating hashtag dynamics with networked groups of generative
  agents.
\newblock In \emph{International {{Conference}} on {{Advances}} in {{Social
  Networks Analysis}} and {{Mining}}}, 2025.

\bibitem[Jiang et~al.(2026)Jiang, Chai, Li, Liu, Fok, Dziri, Tsvetkov, Sap, and
  Choi]{jiang2026artificial}
Liwei Jiang, Yuanjun Chai, Margaret Li, Mickel Liu, Raymond Fok, Nouha Dziri,
  Yulia Tsvetkov, Maarten Sap, and Yejin Choi.
\newblock Artificial hivemind: The open-ended homogeneity of language models
  (and beyond).
\newblock In \emph{The Thirty-ninth Annual Conference on Neural Information
  Processing Systems Datasets and Benchmarks Track}, 2026.
\newblock URL \url{https://openreview.net/forum?id=saDOrrnNTz}.

\bibitem[Kong et~al.(2026)Kong, Lai, Piao, and
  Evans]{kong2026multillmsystemsexhibitrobust}
Weiyi Kong, Shiyang Lai, Jinghua Piao, and James Evans.
\newblock Multi-llm systems exhibit robust semantic collapse, 2026.
\newblock URL \url{https://arxiv.org/abs/2605.17193}.

\bibitem[Li et~al.(2024)Li, Zhang, Yu, Fu, and Ye]{liMoreAgentsAll2024a}
Junyou Li, Qin Zhang, Yangbin Yu, Qiang Fu, and Deheng Ye.
\newblock More {{Agents Is All You Need}}.
\newblock https://arxiv.org/abs/2402.05120v2, February 2024.

\bibitem[Li et~al.(2025)Li, Lin, Xia, and
  Jin]{li2025rethinkingmixtureofagentsmixingdifferent}
Wenzhe Li, Yong Lin, Mengzhou Xia, and Chi Jin.
\newblock Rethinking mixture-of-agents: Is mixing different large language
  models beneficial?, 2025.
\newblock URL \url{https://arxiv.org/abs/2502.00674}.

\bibitem[Lin et~al.(2025{\natexlab{a}})Lin, Huang, Ku, Hsu, Shieh, and
  Kang]{linSimulatingSocialInfluence2025}
Hsien-Tsung Lin, Pei-Cing Huang, Chan-Tung Ku, Chan Hsu, Pei-Xuan Shieh, and
  Yihuang Kang.
\newblock Towards {{Simulating Social Influence Dynamics}} with {{LLM-Based
  Multi-Agents}}.
\newblock In \emph{2025 {{IEEE International Conference}} on {{Information
  Reuse}} and {{Integration}} and {{Data Science}} ({{IRI}})}, pp.\  307--312,
  August 2025{\natexlab{a}}.
\newblock \doi{10.1109/IRI66576.2025.00064}.

\bibitem[Lin et~al.(2025{\natexlab{b}})Lin, Ojha, Cai, and
  Chen]{lin2025strategiccollusionllmagents}
Ryan~Y. Lin, Siddhartha Ojha, Kevin Cai, and Maxwell~F. Chen.
\newblock Strategic collusion of llm agents: Market division in multi-commodity
  competitions, 2025{\natexlab{b}}.
\newblock URL \url{https://arxiv.org/abs/2410.00031}.

\bibitem[Lin et~al.(2025{\natexlab{c}})Lin, Chen, Li, Wu, Wu, Chen, Lee, and
  Chen]{linCreativityLLMbasedMultiAgent2025}
Yi-Cheng Lin, Kang-Chieh Chen, Zhe-Yan Li, Tzu-Heng Wu, Tzu-Hsuan Wu, Kuan-Yu
  Chen, Hung-yi Lee, and Yun-Nung Chen.
\newblock Creativity in {{LLM-based Multi-Agent Systems}}: {{A Survey}}.
\newblock In Christos Christodoulopoulos, Tanmoy Chakraborty, Carolyn Rose, and
  Violet Peng (eds.), \emph{Proceedings of the 2025 {{Conference}} on
  {{Empirical Methods}} in {{Natural Language Processing}}}, pp.\
  27584--27607, Suzhou, China, November 2025{\natexlab{c}}. Association for
  Computational Linguistics.
\newblock ISBN 979-8-89176-332-6.
\newblock \doi{10.18653/v1/2025.emnlp-main.1403}.

\bibitem[L{\'o}pez et~al.(2025{\natexlab{a}})L{\'o}pez, Pastor, Aguilera,
  T{\'a}rraga, Chac{\'o}n, {Pastor-Galindo}, and
  {Ruip{\'e}rez-Valiente}]{lopezAgentbasedSimulationOnline2025}
Alejandro~Buitrago L{\'o}pez, Alberto~Ortega Pastor, D.~M. Aguilera,
  Mario~Fern{\'a}ndez T{\'a}rraga, Jes{\'u}s~Verd{\'u} Chac{\'o}n, Javier
  {Pastor-Galindo}, and Jos{\'e}~A. {Ruip{\'e}rez-Valiente}.
\newblock Agent-based simulation of online social networks and disinformation.
\newblock \emph{ArXiv}, abs/2512.22082, 2025{\natexlab{a}}.

\bibitem[L{\'o}pez et~al.(2025{\natexlab{b}})L{\'o}pez, {Pastor-Galindo}, and
  {Ruip{\'e}rez-Valiente}]{lopezSyntheticGenerationOnline2025}
Alejandro~Buitrago L{\'o}pez, Javier {Pastor-Galindo}, and Jos{\'e}~A.
  {Ruip{\'e}rez-Valiente}.
\newblock Synthetic generation of online social networks through homophily.
\newblock \emph{ArXiv}, abs/2509.02762, 2025{\natexlab{b}}.

\bibitem[Maurya et~al.(2025)Maurya, Shukla, Dandekar, Dandekar, and
  Panat]{maurya2025simulatingmisinformationpropagationsocial}
Raj~Gaurav Maurya, Vaibhav Shukla, Raj~Abhijit Dandekar, Rajat Dandekar, and
  Sreedath Panat.
\newblock Simulating misinformation propagation in social networks using large
  language models, 2025.
\newblock URL \url{https://arxiv.org/abs/2511.10384}.

\bibitem[McPherson et~al.(2001)McPherson, {Smith-Lovin}, and
  Cook]{mcphersonBirdsFeatherHomophily2001}
Miller McPherson, Lynn {Smith-Lovin}, and James~M Cook.
\newblock Birds of a {{Feather}}: {{Homophily}} in {{Social Networks}}.
\newblock \emph{Annual Review of Sociology}, 27\penalty0 (1):\penalty0
  415--444, August 2001.
\newblock ISSN 0360-0572, 1545-2115.
\newblock \doi{10.1146/annurev.soc.27.1.415}.

\bibitem[Mehdizadeh \& Hilbert(2025)Mehdizadeh and
  Hilbert]{mehdizadehHomophilyinducedEmergenceBiased2025}
Aliakbar Mehdizadeh and Martin Hilbert.
\newblock Homophily-induced emergence of biased structures in {{LLM-based}}
  multi-agent {{AI}} systems.
\newblock \emph{Social Network Analysis and Mining}, 15, 2025.

\bibitem[Minsky(1986)]{Minsky1986}
Marvin Minsky.
\newblock \emph{The Society of Mind}.
\newblock Simon \& Schuster, New York, NY, 1986.
\newblock ISBN 978-0671607401.

\bibitem[{Mistral AI}(2025)]{mistral2025magistral}
{Mistral AI}.
\newblock Magistral.
\newblock \emph{arXiv preprint arXiv:2506.10910}, 2025.
\newblock URL \url{https://arxiv.org/abs/2506.10910}.

\bibitem[Monroe et~al.(2008)Monroe, Colaresi, and Quinn]{monroe2008fightin}
Burt~L Monroe, Michael~P Colaresi, and Kevin~M Quinn.
\newblock Fightin' words: Lexical feature selection and evaluation for
  identifying the content of political conflict.
\newblock \emph{Political Analysis}, 16\penalty0 (4):\penalty0 372--403, 2008.

\bibitem[Mou et~al.(2024)Mou, Ding, He, Wang, Liang, Zhang, Sun, Lin, Zhou,
  Huang, and Wei]{mouIndividualSocietySurvey2024}
Xinyi Mou, Xuanwen Ding, Qi~He, Liang Wang, Jingcong Liang, Xinnong Zhang, Libo
  Sun, Jiayu Lin, Jie Zhou, Xuanjing Huang, and Zhongyu Wei.
\newblock From {{Individual}} to {{Society}}: {{A Survey}} on {{Social
  Simulation Driven}} by {{Large Language Model-based Agents}}.
\newblock https://arxiv.org/abs/2412.03563v1, December 2024.

\bibitem[M{\"u}nker et~al.(2025)M{\"u}nker, Schwager, and
  Rettinger]{munkerDontTrustGenerative2025}
Simon M{\"u}nker, Nils Schwager, and Achim Rettinger.
\newblock Don't {{Trust Generative Agents}} to {{Mimic Communication}} on
  {{Social Networks Unless You Benchmarked}} their {{Empirical Realism}}.
\newblock \emph{ArXiv}, abs/2506.21974, 2025.

\bibitem[Nudo et~al.(2025)Nudo, Pandolfo, Loru, Samory, Cinelli, and
  Quattrociocchi]{nudoGenerativeExaggerationLLM2025}
Jacopo Nudo, Mario~Edoardo Pandolfo, Edoardo Loru, Mattia Samory, Matteo
  Cinelli, and Walter Quattrociocchi.
\newblock Generative {{Exaggeration}} in {{LLM Social Agents}}:
  {{Consistency}}, {{Bias}}, and {{Toxicity}}.
\newblock \emph{Online Soc. Networks Media}, 51:\penalty0 100344, 2025.

\bibitem[OpenAI(2025)]{openai2025gptoss120bgptoss20bmodel}
OpenAI.
\newblock gpt-oss-120b \& gpt-oss-20b model card, 2025.
\newblock URL \url{https://arxiv.org/abs/2508.10925}.

\bibitem[Papachristou \& Yuan(2024)Papachristou and
  Yuan]{papachristouNetworkFormationDynamics2024}
Marios Papachristou and Yuan Yuan.
\newblock Network formation and dynamics among multi-{{LLMs}}.
\newblock \emph{PNAS Nexus}, 4, 2024.

\bibitem[Park et~al.(2022)Park, Popowski, Cai, Morris, Liang, and
  Bernstein]{parkSocialSimulacraCreating2022}
J.~Park, Lindsay Popowski, Carrie~J. Cai, M.~Morris, Percy Liang, and
  Michael~S. Bernstein.
\newblock Social {{Simulacra}}: {{Creating Populated Prototypes}} for {{Social
  Computing Systems}}.
\newblock \emph{Proceedings of the 35th Annual ACM Symposium on User Interface
  Software and Technology}, 2022.

\bibitem[Park et~al.(2023{\natexlab{a}})Park, O'Brien, Cai, Morris, Liang, and
  Bernstein]{parkGenerativeAgentsInteractive2023}
Joon~Sung Park, Joseph O'Brien, Carrie~Jun Cai, Meredith~Ringel Morris, Percy
  Liang, and Michael~S. Bernstein.
\newblock Generative {{Agents}}: {{Interactive Simulacra}} of {{Human
  Behavior}}.
\newblock In \emph{Proceedings of the 36th {{Annual ACM Symposium}} on {{User
  Interface Software}} and {{Technology}}}, {{UIST}} '23, pp.\  1--22, New
  York, NY, USA, October 2023{\natexlab{a}}. Association for Computing
  Machinery.
\newblock ISBN 979-8-4007-0132-0.
\newblock \doi{10.1145/3586183.3606763}.

\bibitem[Park et~al.(2023{\natexlab{b}})Park, O'Brien, Cai, Morris, Liang, and
  Bernstein]{park2023generativeagentsinteractivesimulacra}
Joon~Sung Park, Joseph~C. O'Brien, Carrie~J. Cai, Meredith~Ringel Morris, Percy
  Liang, and Michael~S. Bernstein.
\newblock Generative agents: Interactive simulacra of human behavior,
  2023{\natexlab{b}}.
\newblock URL \url{https://arxiv.org/abs/2304.03442}.

\bibitem[{Pastor-Galindo} et~al.(2023){Pastor-Galindo}, Nespoli, and
  {Ruip{\'e}rez-Valiente}]{pastor-galindoLargeLanguageModelPoweredAgentBasedFramework2023}
Javier {Pastor-Galindo}, P.~Nespoli, and Jos{\'e}~A. {Ruip{\'e}rez-Valiente}.
\newblock Large-{{Language-Model-Powered Agent-Based Framework}} for
  {{Misinformation}} and {{Disinformation Research}}: {{Opportunities}} and
  {{Open Challenges}}.
\newblock \emph{IEEE Security \& Privacy}, 22:\penalty0 24--36, 2023.

\bibitem[Pham(2025)]{phamSchemingAbilityLLMtoLLM2025}
Thao Pham.
\newblock Scheming {{Ability}} in {{LLM-to-LLM Strategic Interactions}}.
\newblock https://arxiv.org/abs/2510.12826v2, October 2025.

\bibitem[Piao et~al.(2025)Piao, Lu, Gao, Xu, Santos, Li, and
  Evans]{piaoEmergenceHumanlikePolarization2025}
J.~Piao, Zhihong Lu, Chen Gao, Fengli Xu, F.~P. Santos, Yong Li, and James
  Evans.
\newblock Emergence of human-like polarization among large language model
  agents.
\newblock \emph{ArXiv}, abs/2501.05171, 2025.

\bibitem[Piao et~al.(2026)Piao, Yan, Zhang, Li, Yan, Lan, Lu, Zheng, Wang,
  Zhou, Gao, Xu, Zhang, Rong, Su, and
  Li]{piaoAgentSocietyLargeScaleSimulation2026}
Jinghua Piao, Yuwei Yan, Jun Zhang, Nian Li, Junbo Yan, Xiaochong Lan, Zhihong
  Lu, Zhiheng Zheng, Jing~Yi Wang, Di~Zhou, Chen Gao, Fengli Xu, Fang Zhang,
  Ke~Rong, Jun Su, and Yong Li.
\newblock {{AgentSociety}}: {{Large-Scale Simulation}} of {{LLM-Driven
  Generative Agents Advances Understanding}} of {{Human Behaviors}} and
  {{Society}}, April 2026.

\bibitem[Plackett(1975)]{plackettAnalysisPermutations1975}
R.~L. Plackett.
\newblock The analysis of permutations.
\newblock \emph{Journal of the Royal Statistical Society. Series C (Applied
  Statistics)}, 24\penalty0 (2):\penalty0 193--202, 1975.
\newblock \doi{10.2307/2346567}.

\bibitem[Potter et~al.(2026)Potter, Crispino, Siu, Wang, and
  Song]{potter2026peerpreservationfrontiermodels}
Yujin Potter, Nicholas Crispino, Vincent Siu, Chenguang Wang, and Dawn Song.
\newblock Peer-preservation in frontier models, 2026.
\newblock URL \url{https://arxiv.org/abs/2604.19784}.

\bibitem[{Qwen Team}(2025)]{qwen3technicalreport}
{Qwen Team}.
\newblock Qwen3 technical report, 2025.
\newblock URL \url{https://arxiv.org/abs/2505.09388}.

\bibitem[Salihefendic(2010)]{redditHotScore}
Amir Salihefendic.
\newblock Reddit's empire of nerds: How a group of programmers made the front
  page of the internet.
\newblock
  \url{https://medium.com/hacking-and-gonzo/how-reddit-ranking-algorithms-work-ef111e33d0d9},
  2010.
\newblock Hot-score formula description.

\bibitem[Surve et~al.(2023)Surve, Rathod, Surana, Malpani, Shamraj, Sankepally,
  Jain, and Mehta]{surveMultiagentSimulatorsSocial2023}
Aditya Surve, Archit Rathod, Mokshit Surana, Gautam Malpani, Aneesh Shamraj,
  Sainath~Reddy Sankepally, Raghav Jain, and Swapneel Mehta.
\newblock Multiagent {{Simulators}} for {{Social Networks}}.
\newblock \emph{ArXiv}, abs/2311.14712, 2023.

\bibitem[Syrnikov et~al.(2026)Syrnikov, Pierucci, Galisai, Prandi, Bisconti,
  Giarrusso, Sorokoletova, Suriani, and
  Nardi]{syrnikov2026institutionalaigoverningllm}
Marcantonio~Bracale Syrnikov, Federico Pierucci, Marcello Galisai, Matteo
  Prandi, Piercosma Bisconti, Francesco Giarrusso, Olga Sorokoletova, Vincenzo
  Suriani, and Daniele Nardi.
\newblock Institutional ai: Governing llm collusion in multi-agent cournot
  markets via public governance graphs, 2026.
\newblock URL \url{https://arxiv.org/abs/2601.11369}.

\bibitem[Truong et~al.(2024)Truong, Lou, Flammini, and
  Menczer]{truong2024quantifyingvulnerabilitiesonlinepublic}
Bao~Tran Truong, Xiaodan Lou, Alessandro Flammini, and Filippo Menczer.
\newblock Quantifying the vulnerabilities of the online public square to
  adversarial manipulation tactics, 2024.
\newblock URL \url{https://arxiv.org/abs/1907.06130}.

\bibitem[Vezhnevets et~al.(2023)Vezhnevets, Agapiou, Aharon, Ziv, Matyas,
  {Du{\'e}{\~n}ez-Guzm{\'a}n}, Cunningham, Osindero, Karmon, and
  Leibo]{vezhnevetsGenerativeAgentbasedModeling2023}
Alexander~Sasha Vezhnevets, John~P. Agapiou, Avia Aharon, Ron Ziv, Jayd Matyas,
  Edgar~A. {Du{\'e}{\~n}ez-Guzm{\'a}n}, William~A. Cunningham, Simon Osindero,
  Danny Karmon, and Joel~Z. Leibo.
\newblock Generative agent-based modeling with actions grounded in physical,
  social, or digital space using {{Concordia}}.
\newblock https://arxiv.org/abs/2312.03664v2, December 2023.

\bibitem[Wang et~al.(2024)Wang, Wang, Athiwaratkun, Zhang, and
  Zou]{wang2024mixtureofagentsenhanceslargelanguage}
Junlin Wang, Jue Wang, Ben Athiwaratkun, Ce~Zhang, and James Zou.
\newblock Mixture-of-agents enhances large language model capabilities, 2024.
\newblock URL \url{https://arxiv.org/abs/2406.04692}.

\bibitem[Weckbecker et~al.(2026)Weckbecker, Müller, Hagag, and
  Mulet]{weckbecker2026thoughtvirusviralmisalignment}
Moritz Weckbecker, Jonas Müller, Ben Hagag, and Michael Mulet.
\newblock Thought virus: Viral misalignment via subliminal prompting in
  multi-agent systems, 2026.
\newblock URL \url{https://arxiv.org/abs/2603.00131}.

\bibitem[Willis et~al.(2026)Willis, Zhao, Du, and
  Leibo]{willisEvaluatingCollectiveBehaviour2026}
Richard Willis, Jianing Zhao, Yali Du, and Joel~Z. Leibo.
\newblock Evaluating {{Collective Behaviour}} of {{Hundreds}} of {{LLM
  Agents}}.
\newblock https://arxiv.org/abs/2602.16662v1, February 2026.

\bibitem[Yang et~al.(2024)Yang, Zhang, Zheng, Jiang, Gan, Wang, Ling, Chen, Ma,
  Dong, Gupta, Hu, Yin, Li, Jia, Wang, Ghanem, Lu, Ouyang, Qiao, Torr, and
  Shao]{yangOASISOpenAgent2024}
Ziyi Yang, Zaibin Zhang, Zirui Zheng, Yuxian Jiang, Ziyue Gan, Zhiyu Wang,
  Zijian Ling, Jinsong Chen, Martz Ma, Bowen Dong, Prateek Gupta, Shuyue Hu,
  Zhenfei Yin, G.~Li, Xu~Jia, Lijun Wang, Bernard Ghanem, Huchuan Lu, Wanli
  Ouyang, Yu~Qiao, Philip Torr, and Jing Shao.
\newblock {{OASIS}}: {{Open Agent Social Interaction Simulations}} with {{One
  Million Agents}}.
\newblock \emph{ArXiv}, abs/2411.11581, 2024.

\bibitem[Ye et~al.(2025)Ye, Liu, Wu, Pang, Yin, Bai, and
  Chen]{yeXMASBuildingMultiAgent2025}
Rui Ye, Xiangrui Liu, Qimin Wu, Xianghe Pang, Zhenfei Yin, Lei Bai, and Siheng
  Chen.
\newblock X-{{MAS}}: {{Towards Building Multi-Agent Systems}} with
  {{Heterogeneous LLMs}}.
\newblock https://arxiv.org/abs/2505.16997v1, May 2025.

\bibitem[Zerhoudi et~al.(2026)Zerhoudi, Dastidar, Klement, Romazanov,
  Einwiller, Dang, Dinzinger, Granitzer, {Hautli-Janisz}, Katzenbeisser,
  Lemmerich, and Mitrovic]{zerhoudiFormFunctionAgent2026}
Saber Zerhoudi, Kanishka~Ghosh Dastidar, Felix Klement, Artur Romazanov,
  Andreas Einwiller, Dang~H. Dang, Michael Dinzinger, Michael Granitzer,
  Annette {Hautli-Janisz}, Stefan Katzenbeisser, Florian Lemmerich, and Jelena
  Mitrovic.
\newblock Form {{Without Function}}: {{Agent Social Behavior}} in the
  {{Moltbook Network}}, March 2026.

\bibitem[Zhang et~al.(2025)Zhang, Lin, Mou, Yang, Liu, Sun, Lyu, Yang, Qi,
  Chen, Li, Yan, Hu, Chen, Wang, Huang, Luo, Tang, Wu, Zhou, and
  Wei]{zhangSocioVerseWorldModel2025}
Xinnong Zhang, Jiayu Lin, Xinyi Mou, Shiyue Yang, Xiawei Liu, Libo Sun, Hanjia
  Lyu, Yihang Yang, Weihong Qi, Yue Chen, Guanying Li, Ling Yan, Yao Hu, Siming
  Chen, Yu~Wang, Xuanjing Huang, Jiebo Luo, Shiping Tang, Libo Wu, Baohua Zhou,
  and Zhongyu Wei.
\newblock {{SocioVerse}}: {{A World Model}} for {{Social Simulation Powered}}
  by {{LLM Agents}} and {{A Pool}} of 10 {{Million Real-World Users}}.
\newblock https://arxiv.org/abs/2504.10157v3, April 2025.

\bibitem[Zhang et~al.(2026)Zhang, Mei, Liu, Wang, Metaxas, Wang, Hamm, and
  Ge]{zhangAgentsWildSafety2026}
Yunbei Zhang, Kai Mei, Ming Liu, Janet Wang, Dimitris~N. Metaxas, Xiao Wang,
  Jihun Hamm, and Yingqiang Ge.
\newblock Agents in the {{Wild}}: {{Safety}}, {{Society}}, and the {{Illusion}}
  of {{Sociality}} on {{Moltbook}}, February 2026.

\bibitem[Zhou et~al.(2026)Zhou, tse Huang, Zhou, Lam, Wang, Zhu, Wang, and
  Sap]{zhou2026pimmurprinciplesensuringvalidity}
Jiaxu Zhou, Jen tse Huang, Xuhui Zhou, Man~Ho Lam, Xintao Wang, Hao Zhu,
  Wenxuan Wang, and Maarten Sap.
\newblock The pimmur principles: Ensuring validity in collective behavior of
  llm societies, 2026.
\newblock URL \url{https://arxiv.org/abs/2509.18052}.

\bibitem[Zhou et~al.(2023)Zhou, Zhu, Mathur, Zhang, Yu, Qi, Morency, Bisk,
  Fried, Neubig, and Sap]{zhouSOTOPIAInteractiveEvaluation2023}
Xuhui Zhou, Hao Zhu, Leena Mathur, Ruohong Zhang, Haofei Yu, Zhengyang Qi,
  Louis-Philippe Morency, Yonatan Bisk, Daniel Fried, Graham Neubig, and
  Maarten Sap.
\newblock {{SOTOPIA}}: {{Interactive Evaluation}} for {{Social Intelligence}}
  in {{Language Agents}}.
\newblock In \emph{The {{Twelfth International Conference}} on {{Learning
  Representations}}}, October 2023.

\bibitem[Zhou et~al.(2025)Zhou, Su, Feng, Zhou, Huang, Kao, Lynch, Volkova, Wu,
  Woolley, Zhu, and Sap]{zhouSOTOPIAS4UserfriendlySystem2025}
Xuhui Zhou, Zhe Su, Sophie Feng, Jiaxu Zhou, Jen-tse Huang, Hsien-Te Kao,
  Spencer Lynch, Svitlana Volkova, Tongshuang Wu, Anita Woolley, Hao Zhu, and
  Maarten Sap.
\newblock {{SOTOPIA-S4}}: A user-friendly system for flexible, customizable,
  and large-scale social simulation.
\newblock In Nouha Dziri, Sean~(Xiang) Ren, and Shizhe Diao (eds.),
  \emph{Proceedings of the 2025 {{Conference}} of the {{Nations}} of the
  {{Americas Chapter}} of the {{Association}} for {{Computational
  Linguistics}}: {{Human Language Technologies}} ({{System Demonstrations}})},
  pp.\  350--360, Albuquerque, New Mexico, April 2025. Association for
  Computational Linguistics.
\newblock ISBN 979-8-89176-191-9.
\newblock \doi{10.18653/v1/2025.naacl-demo.30}.

\bibitem[Zhu et~al.(2025)Zhu, He, Haq, Tyson, and
  Hui]{zhuCharacterizingLLMdrivenSocial2025}
Yiming Zhu, Yupeng He, Ehsan-Ul Haq, Gareth Tyson, and Pan Hui.
\newblock Characterizing {{LLM-driven Social Network}}: {{The Chirper}}.ai
  {{Case}}.
\newblock \emph{ArXiv}, abs/2504.10286, 2025.

\end{thebibliography}
